\documentclass[12pt]{extarticle}
\pdfoutput=1

\usepackage[T1]{fontenc}
\usepackage{mathtools}
\usepackage{amssymb}
\usepackage{braket}
\usepackage{tensor}
\usepackage{slashed}
\usepackage{dsfont}
\usepackage{graphicx}
\usepackage{array}
\usepackage{float}
\usepackage{xcolor}
\usepackage{enumitem}
\usepackage{setspace}
\usepackage{geometry}
\usepackage{tikz}
\usetikzlibrary{
    shapes.callouts,
    positioning,
    calc
}
\usepackage[
    colorlinks=true,
    linkcolor=blue,
    citecolor=red,
    urlcolor=blue
]{hyperref}
\numberwithin{equation}{section}
\allowdisplaybreaks[2]

\newcommand{\tr}{\operatorname{tr}}
\newcommand{\pd}{\partial}

\newcommand{\bz}{\mathbb{Z}}

\title{Localization and Abelianization of Strings \\ on Group Manifolds: \\ The Simply Connected Case
}

\makeatletter
\renewcommand\@date{
  \vspace{-\baselineskip}
  \large\centering
  Yongchao L\"u

  \normalsize
  lychaoaa@gmail.com
}
\makeatother

\begin{document}
\maketitle
\begin{abstract}
We study the localization formula for the partition function of the Wess--Zumino--Witten (WZW) model on the torus for compact, connected, and simply connected simple Lie groups. We identify a missing factor in the original localization treatment of Murthy and Witten and show that it agrees with the result obtained from the Hamiltonian formulation. We trace its origin to the abelianization of the Wess–Zumino amplitude, which reduces to a flat Kalb–Ramond $B$–field holonomy in a Narain lattice CFT associated with the maximal torus. This establishes a direct relation between the localization result and the lattice description of the corresponding abelian theory. We further analyze the point–particle limit of the torus partition function, in which the WZW model reduces to quantum mechanics on the group manifold, and reproduce Frenkel’s heat-kernel trace formula.

\end{abstract}

\tableofcontents

\section{Introduction}

Wess--Zumino--Witten (WZW) models provide important examples of exactly solvable interacting two--dimensional conformal field theories (CFT) describing string propagation on group manifolds~\cite{Witten1983WZW}. In the canonical Hamiltonian quantization approach, their mathematical structure is well understood from the algebraic perspective of affine Kac--Moody representation theory~\cite{KnizhnikZamolodchikov1984WZW}. In particular, on a toroidal worldsheet, the partition function is expressed in terms of characters of integrable affine Kac--Moody representations~\cite{gepnerWitten86}, which are given explicitly by the Weyl--Kac character formula~\cite{kac3rd}.

On the other hand, the WZW model admits a Lagrangian formulation as a nonlinear sigma model whose fields are maps from the worldsheet into the compact connected simple group manifold, supplemented by the topological Wess–Zumino (WZ) term~\cite{Wesszumino1971}. In this work, we focus on the \emph{simply connected} case, thereby avoiding additional complications associated with the global topology of the Lie group. Although the theory is exactly solvable as a 2d CFT, a direct evaluation of its path integral remains highly nontrivial. A novel approach initiated by Murthy and Witten utilizes the relation between the WZW model and its supersymmetric extension~\cite{murthyW25}. The supersymmetric WZW (SWZW) model can be represented as a decoupled system of a bosonic WZW model and free Majorana fermions valued in the Lie algebra~\cite{DiVecchiaKnizhnikPR1984SWZW}. In the Hamiltonian approach, with periodic boundary conditions for the fermions and with the neutral fermion zero modes lifted, the contribution of the adjoint Majorana--Weyl fermions cancels the denominator in the Weyl--Kac character formula, so that the SWZW torus partition function isolates precisely the numerator of the WZW partition function.

Murthy and Witten then show that the path integral of the SWZW model can be evaluated exactly using supersymmetric localization, by choosing a suitable real supercharge $Q$ and a corresponding localization term. More precisely, they impose periodic boundary conditions for the fermions and insert fermion zero modes to obtain a non-vanishing path integral. Despite the strong constraints imposed by these zero--mode insertions, a suitable localization term can still be constructed. The localization argument then expresses the path integral as a sum of one-loop contributions around the critical points of the $Q$--exact term. In particular, the localization equations abelianize the theory, reducing the localization locus to isolated configurations in the normalizer of the maximal torus of the Lie group, labelled by the cocharacter lattice and the Weyl group. This leads to a remarkable localization formula for the SWZW partition function, and hence for the numerator of the WZW partition function.

A careful examination of the localization computation in~\cite{murthyW25} reveals a global contribution from the Wess–Zumino (WZ) amplitude that was missing in their original treatment of the classical contribution. Although the Cartan three-form on the Lie group vanishes upon restriction to the maximal torus, the WZ amplitude evaluated on a maximal-torus-valued field on the toroidal worldsheet can remain nontrivial, since its definition involves an extension of the field into the full Lie group. More invariantly, the WZ amplitude is the holonomy of the basic bundle gerbe on the Lie group~\cite{GawedzkiReis2002braneGerbe}. Upon restriction to the maximal torus, this gerbe becomes a flat abelian gerbe whose curvature vanishes while its holonomy can remain nontrivial. This residual holonomy produces an additional phase in the localization formula, which we show precisely matches the phase obtained independently from the diagonal modular invariant of affine characters in the Hamiltonian formulation~\cite{gepnerWitten86}. This missing WZ contribution and its associated phase were communicated to Murthy and Witten in an earlier draft of the present work and were subsequently incorporated, with acknowledgment, into the revised and published versions of~\cite{murthyW25}.

The global origin of the additional phase also admits a useful interpretation in the abelianized theory. Upon abelianization, the semiclassical configurations are effectively captured by a sigma model on the maximal torus, equivalently an abelian WZW model, which admits a natural description as a Narain toroidal CFT~\cite{Narain:1986am}. The SWZW partition function can therefore be viewed as a Weyl--group sum of supersymmetric Narain partition functions, with each contribution described by a Siegel--Narain theta function. In this description, the additional phase is naturally identified with the topological contribution of the flat Kalb--Ramond $B$-field coupling. This flat $B$-field encodes the flat abelian gerbe obtained by restricting the basic gerbe to the maximal torus, thereby confirming the geometric interpretation above. This provides a complementary interpretation of the global WZ holonomy and makes explicit the lattice structure underlying the localization formula.

Through circle reduction to one dimension, the WZW and SWZW models reduce to quantum mechanics on the group manifold and its supersymmetric extension~\cite{choiT25}. Although the WZ amplitude becomes trivial in the one-dimensional reduction, the partition functions retain rich information about the geometry and representation theory of the compact Lie group, encoded in the corresponding heat-kernel trace formulas. On the supersymmetric side, the partition function also admits a localization description, which takes the form of an affine Weyl--group sum. The equivalence between the heat-kernel and localization expressions was mathematically established by Frenkel in his study of the orbit theory of affine Kac--Moody algebras~\cite{Frenkel1984}. The quantum-mechanical limit therefore provides an independent check of the localization formula obtained in the two-dimensional theory. Moreover, a suitable specialization of Frenkel’s formula yields a heat-kernel representation of the affine numerator in the Weyl--Kac formula, in agreement with the heat-equation constraint implied by the Ward identities of the affine current algebra~\cite{Bernard:1987df}.

The paper is organized as follows. In Section~\ref{sec:hamiltonian}, we derive the localization formula from the torus partition function in the Hamiltonian formulation and identify the additional phase absent in the original localization treatment. In Section~\ref{sec:localization}, we show that this phase originates from the WZ amplitude evaluated on the abelianized localizing solutions.  In Section~\ref{sec:narain}, we interpret this holonomy in terms of the flat $B$–field that emerges upon abelianization, relating the SWZW partition function to a Weyl--group sum of supersymmetric Narain partition functions and exhibiting the underlying Narain lattice structure. In Section~\ref{sec:frenkel}, we study the point--particle limit, recover Frenkel’s heat-kernel trace formula, and derive a heat-equation form of the affine numerator formula from a suitable specialization. Finally, Section~\ref{sec:summary} summarizes our results and discusses further implications and directions. Several technical details are collected in the appendices. Appendix~\ref{apx:gramMatrix} summarizes the parity properties of the simple-coroot Gram matrices. Appendix~\ref{apx:bundleWZW} discusses the localization equations and their formulation in terms of holomorphic adjoint bundles. Appendix~\ref{apx:WZ&Bfield} develops the relation between the WZ amplitude and the flat $B$--field amplitude, including the reduction of the PW cocycle on the maximal torus. Appendix~\ref{apx:WZKRgauging} discusses the gauging of the WZ term and the corresponding $B$--field coupling. Appendix~\ref{apx:solidTorus} gives an explicit solid-torus extension of the torus-valued classical configurations and a direct evaluation of their WZ amplitude.

\section{From Hamiltonian formalism to localization formula} \label{sec:hamiltonian}

In this section, we derive the localization formula for the torus partition function of the SWZW model from its Hamiltonian description. The model consists of a WZW sector coupled to free Lie-algebra-valued Majorana fermions, and its partition function accordingly factorizes into bosonic and fermionic contributions. With periodic boundary conditions, the fermionic contribution cancels the Weyl–Kac denominator, leaving the SWZW partition function as a sesquilinear sum of Weyl–Kac numerators.\footnote{Related aspects of the $SU(N)$ case were considered in~\cite{Zhao:2025hen}.} Rewriting these numerators in terms of classical theta functions and performing Poisson resummation then yields the localization formula of Murthy and Witten~\cite{murthyW25}, supplemented by a new phase factor that is generically nontrivial.

We first fix some relevant notation. Let $\mathfrak g$ be a simple Lie algebra with Cartan subalgebra $\mathfrak h$, Weyl group $W$, weight lattice $P$, and coroot lattice $Q^\vee$. Let $\theta$ denote the highest root of $\mathfrak g$. We use the invariant bilinear form $(\cdot ,\cdot)$, normalized by $(\theta,\theta)=2$, to identify $\mathfrak h$ with its dual $\mathfrak h^\ast$. Let $\rho$ denote the Weyl vector, namely half the sum of the positive roots, and define the dual Coxeter number by $h^\vee=1+(\rho,\theta)$.
The set of dominant weights is $P_+=P\cap C_0$, where $C_0\subset\mathfrak h$ is the closed fundamental Weyl chamber. We also denote by $P_{++}=P\cap\mathring C_0$ the set of regular dominant weights, where $\mathring C_0$ is the interior of the fundamental Weyl chamber.
The untwisted affine Kac–Moody algebra at level $k$ is denoted by $\hat{\mathfrak g}^{(1)}_k$ and is obtained as the central extension of the loop algebra of $\mathfrak g$ at level $k$. Its integrable highest-weight representations are labelled by
$P_+^k=\{\lambda\in P_+\mid(\lambda,\theta)\leq k\}$.
Writing $\kappa=k+h^\vee$, we define the corresponding shifted alcove
$P_{++}^{\kappa}
=\{\lambda\in P_{++}\mid(\lambda,\theta) < \kappa\}$.

\subsection{The WZW and SWZW partition functions}

Now consider the WZW model for a compact \emph{simply connected} simple Lie group $G$ with Lie algebra $\mathfrak{g}$, with an integer-valued level $k$. Its Hamiltonian formulation is based on the left- and right-moving affine current algebras
\begin{align}
    \hat{\mathfrak g}^{(1)}_k
\times
\hat{\mathfrak g}^{(1)}_k.
\end{align}
The Hilbert space decomposes into sectors labeled by pairs $(\lambda,\lambda)$ where $\lambda \in P^k_+$~\cite{gepnerWitten86}.
The affine character $\chi_{\lambda,k}$ for $\lambda\in P^k_+$ is given by the Weyl–Kac formula~\cite{kac3rd}
\begin{align}
    \chi_{\lambda,k} = \frac{N_{\lambda+\rho, \kappa}}{N_{\rho,h^\vee}},
\end{align}
where the Weyl–Kac numerator is
\begin{align}
\label{eq:WeylKacNum}
    N_{\lambda,\kappa}
=
\sum_{\sigma\in W}
\det(\sigma)\,
\Theta_{\sigma(\lambda),\kappa}.
\end{align}
Here $\Theta_{\lambda,\kappa}$ is the level-$\kappa$ theta function associated with the coroot lattice,
\begin{align}
\label{eq:classicalTheta}
    \Theta_{\lambda,\kappa}(\tau,u)
=
\sum_{\gamma\in Q^\vee}
\exp\{\frac{\pi i \tau}{\kappa}(\lambda+\kappa\gamma)^2 + 2 \pi i (\lambda+\kappa\gamma,u)\}
.
\end{align}
Thus, the nontrivial dependence on the label $\lambda$ is entirely encoded in the Weyl–Kac numerator, while the denominator is universal, independent of $\lambda$.

The torus partition function is obtained from a trace over the
Hilbert space in the presence of complexified background holonomies.
It can be written in a holomorphically factorized form, up to a
non-holomorphic prefactor:
\begin{align}
    Z_{{\rm WZW}} = C_k \sum_{\lambda \in P^k_+} \chi_{\lambda,k}(\tau,u) \overline{\chi_{\lambda,k}(\tau,v)},
\end{align}
where $\tau = \tau_1 + i \tau_2$ with $\tau_2 > 0$ is the modular parameter of the torus, and $u, v \in \mathfrak{h}_{\mathbb{C}}$ parametrize the background holonomies for the $G^L \times G^R$ global symmetry, and~\cite{murthyW25}
\begin{align}
    C_k = \exp\{ \frac{\pi ik}{\tau_2}\left((u, \mathrm{Im}\,u ) + (\bar{v}, \mathrm{Im}\,\bar v )\right)\}.
\end{align}
The prefactor $C_k$ is the equivariant anomaly counterterm associated with the $G^L \times G^R$ global symmetry. Its modular transformation compensates the anomalous phase of the affine characters in the presence of background holonomies, ensuring the modular invariance of the partition function.

As a compact notation, we write
\begin{align}
    \bar{f}_{\lambda}
    &\equiv
    \overline{f_{\lambda}(\tau,v)}=
    f_{\lambda}(-\bar{\tau},-\bar{v})
    ,
\end{align}
where $f_{\lambda}$ denotes any of
$\chi_{\lambda,k}$, $N_{\lambda,\kappa}$, or
$\Theta_{\lambda,\kappa}$.

We also consider the partition function of Majorana fermions in the adjoint representation of $G$, with periodic boundary conditions on $T^2$. It is given by
\begin{align}
    Z_F = 2^r C_{h^\vee}N_{\rho,h^\vee} \bar{N}_{\rho,h^\vee},
\end{align}
where
\begin{align}
    C_{h^\vee} = \exp\{ \frac{\pi ih^\vee}{\tau_2}\left((u, \mathrm{Im}\,u ) + (\bar{v}, \mathrm{Im}\, \bar v )\right)\},
\end{align}
and the factor $2^r$ with $r = \text{rank}(G)$ arises from the quantization of the neutral Majorana zero modes associated with the Cartan subalgebra.

The SWZW model is obtained by combining the WZW model with Majorana fermions in the adjoint representation, and the partition function factorizes as
\begin{align} \label{eq:swzw1}
    Z_{\text{SWZW}}=Z_F\, Z_{\text{WZW}}.
\end{align}
For periodic boundary conditions on the fermions, the factor
$N_{\rho,h^\vee}\bar N_{\rho,h^\vee}$ cancels the  denominator
of the WZW partition function, while the anomaly prefactors combine according to 
\begin{align}
    C_{\kappa} = C_{k} C_{h^\vee} = \exp\{ \frac{\pi i \kappa}{\tau_2}\left((u, \mathrm{Im}\,u ) + (\bar{v}, \mathrm{Im}\, \bar v )\right)\} .
\end{align}
The remaining partition function therefore depends only on the Weyl--Kac numerators at the shifted level $\kappa=k+h^\vee$,
\begin{align} \label{eq:swzw2}
    Z_{\text{SWZW}} =   2^r C_{\kappa} \sum_{\lambda \in P^k_+} N_{\lambda+\rho,\kappa}(\tau,u) \overline{N_{\lambda+\rho,\kappa}(\tau,v)} .
\end{align}
The shift $\lambda\mapsto\lambda+\rho$ identifies $P_+^k$ with $P_{++}^{\kappa}$, hence
\begin{align}
    Z_{\rm SWZW} =  2^rC_\kappa  \sum_{\lambda \in P^\kappa_{++}} N_{\lambda ,\kappa} \bar{N}_{\lambda,\kappa} .
\end{align}
The Weyl--Kac numerator is antisymmetric under the Weyl group, $N_{\sigma(\lambda),\kappa}=\det(\sigma)N_{\lambda,\kappa}$, and therefore vanishes whenever its label lies on a wall of the affine Weyl alcove. We may consequently extend the sum from $P_{++}^{\kappa}$ to $P_+^\kappa$, obtaining
\begin{align}
   Z_{\rm SWZW} = 2^rC_\kappa \sum_{\lambda \in P^\kappa_{+}} N_{\lambda ,\kappa} \bar{N}_{\lambda,\kappa}.
\end{align}
Expanding $\bar{N}_{\lambda ,\kappa} $ and using the Weyl anti-symmetry of $N_{\lambda,\kappa}$, we obtain
\begin{align}
     Z_{\rm SWZW} 
    = 2^rC_\kappa 
    \sum_{\lambda \in P^\kappa_+}
    \sum_{\sigma\in W}
    N_{\sigma(\lambda),\kappa}\,
    \bar{\Theta}_{\sigma(\lambda),\kappa} .
\end{align}
The Weyl action unfolds the sum over $P^\kappa_+$ into a sum over $P/\kappa Q^\vee$, giving
\begin{align}
     Z_{\rm SWZW} 
    =
    2^rC_\kappa  \sum_{\lambda\in P/\kappa Q^\vee}
    N_{\lambda,\kappa}\,
    \bar{\Theta}_{\lambda,\kappa}.
\end{align}
Expanding $N_{\lambda,\kappa}$ then gives
\begin{align}
    Z_{\rm SWZW} = 2^rC_\kappa  \sum_{\sigma\in W}\det(\sigma)
    \sum_{\lambda\in P/\kappa Q^\vee}
    \Theta_{\sigma(\lambda),\kappa}\,
    \bar{\Theta}_{\lambda,\kappa},
\end{align}
Using the Weyl covariance of the theta functions,
\begin{align}
\Theta_{\sigma(\lambda),\kappa}(\tau, u) = \Theta_{\lambda,\kappa}(\tau, \sigma(u)), \quad \sigma \in W,
\end{align}
we obtain
\begin{align} 
\label{eq:swzwTheta}
     Z_{\rm SWZW}  = 2^rC_\kappa  \sum_{\sigma\in W}\det(\sigma)
    \sum_{\lambda\in P/\kappa Q^\vee}
    \Theta_{\lambda,\kappa}(\tau, \sigma(u) )\,
   \overline{ \Theta_{\lambda,\kappa}(\tau, v) }    .
\end{align}
It is therefore useful to introduce the diagonal theta-function combination
\begin{align}
\label{eq:McZ}
    \mathcal{Z} = 2^rC_\kappa \sum_{\lambda\in P/\kappa Q^\vee}
    \Theta_{\lambda,\kappa}\,
    \bar{\Theta}_{\lambda,\kappa} ,
\end{align}
then
\begin{align}
\label{eq:ZswzwinMcZ}
    Z_{\rm SWZW} =  \sum_{\sigma \in W} \det(\sigma) \mathcal{Z}^\sigma,
\end{align}
where $\mathcal{Z}^\sigma(\tau,u,v) = \mathcal{Z}(\tau, \sigma(u),v)$.
Notice that $C_\kappa$ is Weyl invariant, so $C_\kappa^\sigma=C_\kappa$. 
The resulting $\mathcal Z$ is therefore the diagonal modular-covariant combination of the level-$\kappa$ theta functions associated with the finite abelian group $P/\kappa Q^\vee$. Consistently, $\mathcal Z$, $Z_{\rm SWZW}$, and $Z_{\rm F}$ all carry modular weight $(r/2,r/2)$, with the same modular behavior as $({\rm Im}\tau)^{-r/2}$, or equivalently $|\eta(\tau)|^{2r}$. They are thus invariant under $T$ and acquire a factor $|\tau|^r$ under $S$. The SWZW partition function is then obtained from $\mathcal Z$ by Weyl antisymmetrization.

\subsection{The localization formula}
The periodicity of the theta functions allows the finite sum over $P/\kappa Q^\vee$ to be unfolded into a sum over the full weight lattice $P$,
\begin{align}
       \mathcal{Z} 
      & = 2^r C_\kappa \sum_{\lambda \in P/\kappa Q^\vee} \sum_{\gamma \in Q^\vee } \exp\{\frac{\pi i \tau}{\kappa}(\lambda+\kappa\gamma)^2 + 2 \pi i (\lambda+\kappa\gamma,u)\} \bar{\Theta}_{\lambda + \kappa \gamma, \kappa} \nonumber \\
     & = 2^r C_\kappa \sum_{\lambda \in P}  \exp\{\frac{\pi i \tau}{\kappa}\lambda^2 + 2 \pi i (\lambda,u)\}\bar{\Theta}_{\lambda, \kappa} .
\end{align}
Expanding the antiholomorphic theta function then introduces a winding variable $w\in Q^\vee$, giving
\begin{align}
\label{eq:McZexpanded}
   \mathcal{Z} 
     = 2^r C_\kappa \sum_{w \in Q^\vee }  \sum_{\lambda \in P}  \exp\{\frac{\pi i \tau}{\kappa}\lambda^2 + 2 \pi i (\lambda,u)-\frac{\pi i \bar\tau}{\kappa}(\lambda+\kappa w)^2 -2 \pi i (\lambda+\kappa w,\bar v)\} .
\end{align}

To prepare for Poisson resummation, it is convenient to separate the dependence on the winding variable $w$ from the Gaussian sum over the momentum lattice $P$. We therefore write
\begin{align}
\mathcal{Z}  & = 2^r C_\kappa \sum_{w \in Q^\vee }  g(w)
\sum_{\lambda \in P}  f_w(\lambda) ,
\end{align}
with
\begin{align}
     g(w) &  = \exp\{-\pi i \overline{\tau} \kappa w^2 - 2 \pi i \kappa (w, \bar{v})\}, \\
     f_w(\lambda)  & = \exp\{\frac{\pi i }{\kappa } (\tau - \overline{\tau} ) \lambda^2  - 2 \pi i \overline{\tau} (\lambda, w) + 2 \pi i (\lambda,u - \bar{v}))\}. 
\end{align}
Now we can apply Poisson summation to the lattice $P$ with $Q^\vee$ as its dual. Thus,
\begin{align}
    \sum_{\lambda \in P } f_w(\lambda)  = \mathrm{Vol}(Q^\vee) \sum_{m \in Q^\vee} F_w(m),
\end{align}
where $F_w(m)$ is the Fourier transform on $\mathfrak{h}$,
\begin{align}
    F_w(m ) =   \int_{\mathfrak{h}} e^{2 \pi i (\lambda,m)} f_w(\lambda) d \lambda
    =  \left( \frac{\kappa}{2\tau_2}\right)^{\frac{r}{2}} \exp\{-\frac{\pi \kappa}{2 \tau_2}(m-\overline{\tau}w + u - \bar{v})^2\} .
\end{align}

The Poisson-resummed expression is therefore naturally labelled by two elements $m,w\in Q^\vee$. The variable $w$ originates from the original theta-function winding sum, while $m$ is the dual lattice variable introduced by Poisson resummation. The resulting Gaussian has precisely the structure expected from classical configurations labelled by $Q^\vee$--valued windings along the worldsheet.

Combining $C_\kappa$, $g(w)$, and $F_w(m)$, we obtain
\begin{align}
\mathcal{Z} 
 =
 2^r \mathrm{Vol}(Q^\vee) (\frac{\kappa}{2 \tau_2})^{\frac{r}{2}} \sum_{m,w\in Q^\vee} e^{-\pi i \kappa (m, w)} H_{m,w},
\label{eq:thetaPoisson}
\end{align}
where
\begin{align} \label{eq:Hmw}
   H_{m,w} = \exp \{ -\frac{\pi \kappa}{2 \tau_2}\left( (m - \tau w  + 2 u, m - \bar{\tau}w   - 2 \bar{v})    + 2 (u , \bar{v}) + (u,\bar{u})+(\bar{v},v) \right) \}
\end{align}
The Gaussian part $H_{m,w}$ reproduces the expected classical contribution of the localization formula. In addition, there appears the phase factor
\begin{align}
\label{eq:xiPhase}
   \xi_{m,w} =  e^{-\pi i\kappa(m,w)}, \quad m,w \in Q^\vee.
\end{align}
Unlike the Gaussian contribution, this phase is independent of the metric parameter encoded in $\tau_2$ and depends only on the pairing of the two windings. This suggests a topological origin, which we will identify in Section~\ref{sec:localization} with the WZ amplitude evaluated on the abelianized localizing solutions.

Finally we obtain the localization
formula for the torus partition function of the SWZW model,
\begin{align} \label{eq:SWZWPartFunc}
    Z_{\text{SWZW}} 
     = 2^r \mathrm{Vol}(Q^\vee) (\frac{\kappa}{2 \tau_2})^{\frac{r}{2}} \sum_{\sigma \in W} \det(\sigma) \sum_{m,w\in Q^\vee} \xi_{m,w}  H^\sigma_{m,w} .
\end{align}
with $H^\sigma_{m,w}(\tau, u, v) = H_{m,w}(\tau, \sigma(u),v)$. 

The Hamiltonian derivation thus reproduces the structure of the Murthy–Witten localization formula, including its Weyl-group sum and lattice-labelled Gaussian contributions, but reveals a new phase factor. This phase factor is absent from the localization formula of Murthy and Witten~\cite{murthyW25} and constitutes the main new result of the present derivation.

\subsection{The new phase factor}
We now analyze the properties of the new phase factor $\xi_{m,w}$ appearing in the localization formula~\eqref{eq:SWZWPartFunc}. Expanding in the basis of simple coroots, $m = \sum_i a_i \alpha_i^\vee$, $w = \sum_i b_i \alpha_i^\vee$, with $a_i,b_i\in\mathbb{Z}$, we obtain
\begin{align}
    (m,w ) = \sum_{i,j}  \mathsf{C}_{i,j}\, a_i b_j,
\end{align}
where
\begin{align}
\label{eq:GramMatrix}
\mathsf{C}_{ij} :=
(\alpha_i^\vee,\alpha_j^\vee).
\end{align}
The matrix $\mathsf{C}(G) = (\mathsf{C}_{ij})$ is the Gram matrix of the simple coroots,  
which is often referred to as the \emph{symmetrized} Cartan matrix. For $m,w\in Q^\vee$,
\begin{align}
    (m,w) \in \mathbb{Z} . 
\end{align}
since $\mathsf{C}(G)$ is an integer matrix.  Since $\kappa=k+h^\vee$ is an integer, the phase factor takes values in $\{\pm1\}$ and can be written as
\begin{align} \label{eq:newphase}
    \xi_{m,w} = (-1)^{\kappa(m,w)}.
\end{align}
Thus the phase depends on the parity of the mixed pairing $(m,w)$.

The diagonal entries of $\mathsf{C}(G)$ are always even, so $(w,w)\in2\mathbb Z$ for any $w\in Q^\vee$;  that is, the coroot lattice $Q^\vee$ is even. However, its bilinear pairing need not be even, as its off-diagonal entries are integral but need not be even. For a simple Lie group $G$, $\xi_{m,w}=1$ for all
$m,w\in Q^\vee$ if and only if
\begin{align}
        \kappa(m,w)\in 2\mathbb{Z}
    \qquad
    \text{for all }m,w\in Q^\vee.
\end{align}
This condition is automatically satisfied when $\kappa$ is even. For odd $\kappa$, it is equivalent to $(m,w)\in2\mathbb Z$ for all $m,w\in Q^\vee$.
The latter condition is equivalent to $\mathsf C(G)$ being entrywise even, which holds precisely for Lie groups of type $C_n$, including $C_1=A_1$. Indeed, for $C_n$,
\begin{align}
\mathsf C(C_n)_{ij}
& =
4\delta_{ij}
-2\delta_{i,j+1}
-2\delta_{i,j-1}
-2\delta_{i,n}\delta_{j,n}\, \quad 1\leq i,j\leq n,
\end{align}
or equivalently
\begin{align}
\mathsf C(C_n)
= 
\begin{pmatrix}
4&-2&0&\cdots&0\\
-2&4&-2&\ddots&\vdots\\
0&-2&4&\ddots&0\\
\vdots&\ddots&\ddots&4&-2\\
0&\cdots&0&-2&2
\end{pmatrix}.
\end{align}

For all other simple Lie groups, $\mathsf C(G)$ has at least one odd entry, as shown explicitly in Appendix~\ref{apx:gramMatrix}. Consequently, there exist $m,w\in Q^\vee$ such
that $(m,w)$ is odd. Hence, for odd $\kappa$, the phase factor is nontrivial for some $m,w\in Q^\vee$.

For instance, consider $G=SU(3)$ with even $k$. Since $h^\vee=3$, the shifted level $\kappa=k+h^\vee=k+3$ is odd. Choosing $m=\alpha_1^\vee$ and $w=\alpha_2^\vee$, we have $(\alpha_1^\vee,\alpha_2^\vee)=-1$. Therefore
\begin{align}
        \xi_{\alpha_1^\vee,\alpha_2^\vee}
    =
    (-1)^{
        \kappa
        (\alpha_1^\vee,\alpha_2^\vee)}=
    -1.
\end{align}
This provides the simplest example in which the new phase is genuinely nontrivial.

\section{Localization and abelianization of the WZ amplitude} \label{sec:localization}

The new phase factor motivates a closer examination of Murthy and Witten’s path-integral derivation of the SWZW partition function on the torus. We show that this phase arises from the WZ amplitude evaluated on the localization locus, which abelianizes to the maximal torus. Although the Cartan three-form vanishes upon restriction to the maximal torus, the WZ amplitude need not be trivial. As a gerbe holonomy, it retains a nontrivial PW cocycle on the abelianized localization locus, which reproduces precisely the phase obtained in the previous section.

First, let us fix some notation for the Lie group $G$. Henceforth, $G$ is assumed to be compact, connected, and simply connected. The group manifold $G$ admits an action of $G\times G$ by left and right multiplication,
\begin{align}
g\mapsto U_L\,g\,U_R^{-1}.
\end{align}
We denote the two $G$ factors by $G_L$ and $G_R$, respectively. The Maurer--Cartan forms are the $\mathfrak g$-valued one-forms
\begin{align}
J=g^{-1}dg,
\quad
K=dgg^{-1},
\end{align}
which satisfy the Maurer--Cartan equations
\begin{align}
dJ+J^2=0,
\quad
dK-K^2=0.
\end{align}
Under the $G_L\times G_R$ action, they transform by conjugation,
\begin{align}
J\mapsto U_RJU_R^{-1},
\quad
K\mapsto U_LKU_L^{-1}.
\end{align}
In particular, $J$ is left-invariant, while $K$ is right-invariant. It follows that the Cartan three-form
\begin{align}
\Omega
=\frac{1}{3}\tr (J^3)
=\frac{1}{3}\tr (K^3)
\end{align}
is invariant under the $G_L\times G_R$ action. The Cartan three-form is closed, and its integral over any closed three-cycle in $G$ is an integer multiple of $8\pi^2$.

\subsection{The gauged WZW model}

The WZW model on a torus $T^2$ is a nonlinear sigma model with compact Lie group $G$ as its target space~\cite{Witten1983WZW}. 
The fundamental field is therefore a map $g:T^2\to G$, and the action takes the form
\begin{align}
    I_{\rm WZW}=I_{\rm kin} + i\Gamma,
\end{align}
where $I_{\rm kin}$ is the sigma-model kinetic term and $\Gamma$ is the Wess--Zumino (WZ) functional~\cite{Wesszumino1971}.

The kinetic term is given by
\begin{align}
I_{\rm kin}(g)
=
-\frac{1}{8\pi}
\int_{T^2}
\tr \left[
J\wedge *J
\right]
=
-\frac{i}{4\pi}
\int_{T^2}dz\wedge d\bar z
\tr \left[
g^{-1}\partial_z g
g^{-1}\partial_{\bar z}g
\right].
\end{align}
Henceforth we use the invariant trace $\mathrm{tr}(\cdot\cdot)$ and
the bilinear form $(\cdot,\cdot)$ interchangeably. 

The WZ functional is defined by extending the field $g$ from $T^2$ to a three-manifold $B$ with boundary $T^2$. Specifically, let $\tilde g:B\to G$ be an extension of $g$. Then the WZ functional is given by the integral over $B$ of the pullback of the Cartan three-form on $G$
\begin{align}
\Gamma_{\rm WZ}(\tilde g)
=
\frac{1}{4\pi}\int_B\tilde g^*\Omega
=
\frac{1}{12\pi}
\int_B
\tr \left[
(\tilde g^{-1}d\tilde g)^3
\right],
\end{align}
Thus, $\Gamma(\tilde g)$ is defined modulo $2\pi\mathbb Z$, with the ambiguity arising from different choices of extension.  For the level $k \in \bz$, the WZ amplitude
\begin{align}
A_{T^2}(g)
=
e^{-ik\Gamma(g)}
\end{align}
is independent of the choice of extension. By abuse of notation, we will henceforth write $\Gamma(g)$ without specifying the extension. The relevant properties of the WZ amplitude and its Polyakov–Wiegmann cocycle are reviewed in Appendix~\ref{apx:WZ&Bfield}. 

The WZW action is invariant under the $G^L\times G^R$ action. We introduce background gauge fields $A^L$ and $A^R$ for $G^L\times G^R$ and subsequently restrict to flat Cartan-valued connections on the torus. These backgrounds encode the left- and right-moving Cartan holonomies that enter the partition function. On $T^2$ with the complex coordinate $z=s+\tau t$,
where $s$ and $t$ each have period $2\pi$, we write the background gauge
fields as
\begin{align}
\label{eq:ALAR}
    A^L
    =
    \frac{\bar v\,dz -v\,d\bar z}{\tau-\bar\tau},
    \quad
    A^R
    =
    \frac{\bar u\,dz-u\,d\bar z}{\tau-\bar\tau}.
\end{align}
The corresponding covariant derivative is
\begin{align}
    D g = dg+A^Lg-gA^R.
\end{align}
The sigma-model kinetic term is gauged by replacing $dg$ with $Dg$:
\begin{align}
I_{\rm kin}(g,A^L,A^R)
& =
-\frac{1}{8\pi}
\int_{T^2}
\tr
g^{-1} Dg 
\wedge
*
g^{-1} Dg 
,\\
& = I_{\rm kin}(g)-  \frac{1}{4 \pi } \int_{T^2} \left( A^L \wedge * gdg^{-1} - g^{-1}dg \wedge * A^R - A^L \wedge * g A^R g^{-1}\right)\\
    & \qquad \qquad  - \frac{1}{8 \pi } \int_{T^2}  \left( A^L \wedge * A^L + A^R \wedge *  A^R  \right) , 
\end{align}
The Cartan three-form $\Omega$ admits a $G_L\times G_R$-equivariant extension $\Omega_A$. On the worldsheet, its pullback takes the form
\begin{align}
g^\ast \Omega_A=  g^\ast \Omega -  d\left( A^L \wedge gdg^{-1} - g^{-1}dg \wedge A^R - A^L \wedge g A^R g^{-1}\right).
\end{align}
Accordingly, the gauged WZ functional is
\begin{align}
     \Gamma_{\rm WZ}(g,A^L,A^R)& = \frac{1}{4 \pi } \int_B g^\ast \Omega_A \\
     & = \Gamma_{\rm WZ}(g) - \frac{1}{4 \pi } \int_{T^2} \left( A^L \wedge gdg^{-1} - g^{-1}dg \wedge A^R - A^L \wedge g A^R g^{-1}\right) 
\end{align}
The construction of the equivariant extension of the Cartan three-form and the resulting gauged WZ term is reviewed in Appendix~\ref{apx:WZKRgauging}.
For later convenience, we combine the kinetic and gauging terms into
\begin{align}
I^A_{\rm kin}(g)
&=
I_{\rm kin}(g)
-\frac{i}{2\pi}\int_{T^2}dz\wedge d\bar z\,
\tr\Big(
A_z^L\partial_{\bar z}g\,g^{-1}
-g^{-1}\partial_zg\,A_{\bar z}^R
-A_z^LgA_{\bar z}^Rg^{-1}
\nonumber\\
&\hspace{7.5cm}
+\frac{1}{2}
\big(
A_z^LA_{\bar z}^L
+A_z^RA_{\bar z}^R
\big)
\Big).
\end{align}
Hence, the gauged WZW action can be written as~\cite{Witten1991holomorpFac}
\begin{align}
    I_{\rm WZW}(g, A^L, A^R) & = I_{\rm kin}(g,A^L, A^R) + i \Gamma_{\rm WZ}(g,A^L,A^R)\\
 & = I_{\rm kin}^A(g) + i \Gamma_{\rm WZ}(g).
\end{align}  
Introducing the covariant Maurer--Cartan one-forms
\begin{align}
    \mathcal{J} = g^{-1} Dg = J + g^{-1} A^L g - A^R ,\quad \mathcal{K} = Dg  g^{-1} =K  + A^L -  g A^R g^{-1} ,\quad  
\end{align}
the equations of motion for $g$ are
\begin{align}
    D_{\bar{z}} \mathcal{J}_{z} + F^R_{\bar{z}z} =0,\quad D_{z} \mathcal{K}_{\bar z} - F^L_{z\bar{z}} = 0. 
\end{align}
For flat background gauge fields, these reduce to
\begin{align}
    D_{\bar{z}} \mathcal{J}_{z} =0,\,\quad  D_{z} \mathcal{K}_{\bar z} =0
\end{align}
The partition function of gauged WZW  model at level $k$ on a torus $T^2$ is defined by the path integral over $g$:
\begin{align}
    Z_{\rm WZW}  = \int \mathcal{D} g \, \exp\{- k I_{\rm WZW}(g, A^L, A^R) \} .
\end{align}

\subsection{The gauged SWZW model and localization}

We now supersymmetrize the gauged WZW model by adding Majorana–Weyl fermions $\psi$ and $\tilde\psi$, valued in the adjoint representation. The fermion $\psi$ couples to the right background gauge field $A^R$, while $\tilde\psi$ couples to the left background gauge field $A^L$. With the covariant derivatives
\begin{align}
   D \psi = d \psi + [A^R, \psi ],\quad D \tilde \psi = d \tilde \psi + [A^L, \tilde \psi ].
\end{align}

The fermions couple minimally to the background gauge fields,
\begin{align}
    I_{F}(\psi,A^R ) = -\frac{1}{4 \pi } \int dz \wedge d\bar{z}\, \tr  \psi D_{\bar z}\psi , \quad     I_{F}(\tilde{\psi},A^L) = -\frac{1}{4 \pi } \int dz \wedge d\bar{z}\, \tr  \tilde{\psi} D_{z} \tilde{\psi} .
\end{align}

The total gauged SWZW action is
\begin{align}
    I_{\rm SWZW}(g, \psi, \tilde \psi, A^L,A^R) = I(g,A^L,A^R) + I_{F}(\psi,A^R)+I_{F}(\tilde{\psi},A^L).
\end{align}
The gauged SWZW action is invariant under the supersymmetry transformations
\begin{align}
    Qg
    =
    i\, g \psi, \quad
     Q \psi
    =
    \mathcal{J}_{z} - i   \psi   \psi   , \quad 
     Q\tilde \psi
    = 0  , \quad  Q A^L = 0 =  Q A^R. 
\end{align} 
and
\begin{align}
    \tilde Qg
    =
    i\,\tilde \psi g , \quad 
    \tilde Q\psi
    = 0 , \quad
    \tilde Q\tilde\psi
    =
     \mathcal{K}_{\bar z} + i \tilde \psi \tilde \psi    , \quad \tilde Q A^L = 0 =  \tilde Q A^R. 
\end{align} 
which satisfies $Q^2 = i D_z$ and $\tilde Q^2 = i D_{\bar z}$ offshell.

Throughout this work we impose periodic (Ramond--Ramond) boundary conditions for the fermions. On the flat Cartan backgrounds relevant for localization, the Cartan components are neutral under the background holonomies and therefore always possess constant zero modes. The gauged SWZW partition function at the shifted level $\kappa=k+h^\vee$ is obtained by integrating over $g$, $\psi$, and $\tilde\psi$, with the Cartan fermion zero modes saturated by the insertion
\begin{align}
\Xi =
(\psi_0^1\cdots\psi_0^r)
(\tilde\psi_0^1\cdots\tilde\psi_0^r),
\end{align}
where $\psi_0^i$ and $\tilde\psi_0^i$ denote the constant zero modes of the Cartan components of $\psi(z)$ and $\tilde\psi(\bar z)$, respectively. The normalization of the zero-mode measure requires a choice of real Clifford-module convention. Following~\cite{murthyW25}, we associate to each left-right Majorana zero-mode pair the grading operator
\begin{align}
(-1)^{F_i}
=
i\psi_0^i\tilde\psi_0^i,
\end{align}
so that $\big((-1)^{F_i}\big)^2=1$. The fermion-parity operator on the full Cartan zero-mode sector is therefore
\begin{align}
(-1)^F
=
\prod_{i=1}^r(-1)^{F_i}.
\end{align}
With this choice of zero-mode normalization, the partition function is
\begin{align}
Z_{\rm SWZW}
=
\int
\mathcal{D}g\,\mathcal{D}\psi\,\mathcal{D}\tilde\psi\,
\Xi\,
\exp\{-\kappa I_{\rm SWZW}\}.
\end{align}
For the localization computation we choose the supercharge $Q$. The fermion zero mode insertion constrains the choice of the $Q$-exact deformation term. 
A convenient choice is
\begin{align}
    V = i \int dz \wedge d\bar z (D_{\bar z} \psi)(D_{z} \mathcal{J}_{\bar z})
\end{align}
such that
\begin{align}
    Q V = i \int dz \wedge d\bar z D_{\bar z}(\mathcal{J}_{z} - i \psi \psi ) D_z \mathcal{J}^L_{\bar{z}} - \int dz \wedge d\bar z D_{\bar{z}}  \psi D_z (D_{\bar{z}}  \psi + [\mathcal{J}_{\bar{z}}, \psi])
\end{align}
Then we get the deformed partition function
\begin{align}
Z_V(\lambda)
=
\int \mathcal{D}g \mathcal{D} \psi \mathcal{D} \tilde \psi\,
\, \Xi \,\exp\{-\kappa I_{\rm SWZW}-\lambda QV\}.
\end{align}
The choice of localization term can ensure that  $Z_V(\lambda)$ is independent of $\lambda$.
In the limit $\lambda \to \infty$, the bosonic part of $QV$ is positive definite, and the path integral localizes onto configurations satisfying
\begin{align}
    \psi=0,\qquad
D_{\bar z}\mathcal J_z=0.
\end{align}
In Appendix~\ref{apx:bundleWZW}, we give a bundle-theoretic interpretation of this equation as the condition that $g_{\rm cl}$ induce a holomorphic isomorphism between the two complexified adjoint bundles. On the abelianized localization locus, this implies that $g_{\rm cl}$ can be written as 
\begin{align}
    g_{\rm cl}=g_\sigma\tilde g,
\end{align}
with $\sigma\in W$ and $\tilde g$ taking values in the maximal torus $T$.
On a torus $T^2$,  the maximal torus valued factor can be written as
\begin{align}
    \tilde g =\exp\{i (mt +ws)\}.
\end{align}
Periodicity of $\tilde g$ implies that $m$, $w$ lie in the cocharacter lattice. For the simply connected group this reduces to $m,w\in Q^\vee$.
Hence, the localization solutions are labeled by 
\begin{align}
    (\sigma,m,w)\in W\times Q^\vee\times Q^\vee,
\end{align}
and are given by
\begin{align} \label{eq:localizingSol}
    g_{\rm cl} = g_\sigma e^{i m t + i w s}= 
g_\sigma
\exp\{
\frac{i}{\tau-\bar\tau}
\big(
\bar\alpha z
-\alpha \bar z
\big)
\}, \quad \alpha = (m-\tau w).
\end{align}

Now we consider the expansion around a localizing solution $g_{\rm cl}$,
\begin{align}
g=g_{\rm cl}e^{y/\sqrt{\lambda}},
\end{align}
together with the rescaling $\psi\to \psi/\sqrt{\lambda}$. 
With $(A^L_z)_g=g^{-1}(\pd_z+A^L_z)g$, the corresponding current has the expansion
\begin{align}
   \sqrt{\lambda} \mathcal{J}_z = \sqrt{\lambda} \mathcal{J}_z^{\rm cl} + (\partial_z + (A^L_z)_{g_{\rm cl}})y  + O(\lambda^{-1/2}) \,, 
\end{align}
where $\mathcal{J}_z^{\rm cl} = (A^L_z)_{g_{\rm cl}} - A^R_z$.
Since the localization equation gives $D_{\bar z}\mathcal{J}_z^{\rm cl}=0$, we obtain
\begin{align}
 \sqrt{\lambda}  D_{\bar z} \mathcal{J}_z =   D_{\bar z} (\partial_z + (A^L_z)_{g_{\rm cl}})y + O(\lambda^{-1/2}).
\end{align}
Thus, at quadratic order around the localization locus,
\begin{align}
\kappa I_{\rm SWZW}+\lambda QV
=
\kappa I_{\rm WZW}(g_{\rm cl},A^L, A^R)
+\kappa I_F(\tilde \psi,A^L)
+(QV)^{(2)}
+O(\lambda^{-1/2}).
\end{align}
Here we used $\psi=0$ on the localization locus, while the $\tilde\psi$ fermion is not scaled away.
The quadratic fluctuations are described by
\begin{align}
    (QV)^{(2)} = (QV)^{(2)}_{\rm bos} + (QV)^{(2)}_{\rm fer},
\end{align}
with
\begin{align}
(QV)^{(2)}_{\rm bos}
=
i \int_{T^2}d^2z\,
\tr
\left|
D_{\bar z} (\partial_z + (A^L_z)_{g_{\rm cl}}))y
\right|^2,
\end{align}
and 
\begin{align}
(QV)^{(2)}_{\rm fer}
=
-\int_{T^2}d^2z\,
\tr
D_{\bar z}
\psi
D_z
(\partial_{\bar z}
+(A^L_{\bar z})_{g_{\rm cl}}
) \psi.
\end{align}
Finally, taking $\lambda \to \infty$ and using the parametrization of the localization solutions $g_{\rm cl}$ by $(\sigma,m,w) \in W \times Q^\vee\times Q^\vee$, we obtain
\begin{align}
    Z_{\rm SWZW} = Z(\infty) = \sum_{(\sigma,m,w)} Z^{\rm cl}_{\sigma,m,w} Z^{\text{1-loop}}_{\sigma,m,w},
\end{align}
where 
the classical contribution is
\begin{align}
 Z^{\rm cl}_{\sigma,m,w} & =e^{-\kappa I_{\rm WZW}(g_{\mathrm{cl}},A^L, A^R)} = e^{-\kappa I_{\rm kin}^A(g_{\rm cl}) } A_{\rm WZ}(g_{\rm cl}), \\
    \quad Z^{\text{1-loop}}_{\sigma,m,w} &= \int\mathcal{D} \tilde \psi\,  e^{- \kappa I_F(\tilde \psi, A^L)} \int\mathcal{D} y\mathcal{D}\psi\, e^{-(QV)^{(2)}}.
\end{align}

\subsection{Abelianization of the WZ amplitude}
Since the localization locus is maximal-torus valued, the pullback of the Cartan three-form vanishes identically. This might suggest that the WZ amplitude is trivial on the localization locus, $A_{\rm WZ}(g_{\rm cl})=1$, as effectively assumed in the original treatment of Murthy and Witten. However, the vanishing of the local three-form $\Omega|_T$ does not imply the triviality of the global WZ amplitude. In general, we will find that
\begin{align}
     A_{\rm WZ}(g_{\rm cl})\neq 1.
\end{align}
This nontrivial WZ amplitude provides the additional phase in the corrected localization formula.

We now evaluate the WZ amplitude on the localization locus. Consider an extension of $g_{\rm cl}$ to the solid torus
$D^2\times S_t^1$,
\begin{align} \label{eq:extension}
    \tilde g_{\rm cl} (s,t,r)=g_\sigma f_w(s,r)e^{ i tm},
\end{align}
where $ 0 \leq r \leq 1 $, and the boundary condition is $f_w(s,1)=e^{ i sw}$. The existence of the extension $f_w:D^2\to G$ follows from the simple connectedness of $G$. Indeed, the loop $s\mapsto e^{ i sw}$ in the maximal torus is contractible in $G$, so it extends continuously to $D^2$. An explicit construction of such an extension using the $SU(2)$ subgroups associated with the simple coroots is given in Appendix~\ref{apx:solidTorus}.

Since $g_\sigma$ is constant, it does not affect the Maurer–Cartan form. Defining $ \theta_w=f_w^{-1}df_w$, $\theta_m = i m\,dt$, we have 
\begin{align}
    \tilde g_{\rm cl}^{-1}d\tilde g_{\rm cl} = e^{- i tm}\theta_we^{ i tm}+\theta_m.
\end{align}
Since $\theta_w$ has support only on $D^2$ and $\theta_m $ only along $S_t^1$, the only nonvanishing contribution to $\mathrm{tr}(\tilde g_{\rm cl}^{-1}d\tilde g_{\rm cl})^3$ is $3\tr \theta_m \theta_w^2$. Therefore
\begin{align}
    \frac{1}{12 \pi}\mathrm{tr}(\tilde g_{\rm cl}^{-1}d\tilde g_{\rm cl})^3
=
\frac{1}{4\pi }\,\mathrm{tr}(\theta_m \theta_w^2) =
-\frac{1}{4\pi}
d\,\tr(\theta_m\wedge\theta_w).
\end{align}
where we have used the Maurer–Cartan equation for $\theta_w$, and $d\theta_m=0$.
Therefore, for the chosen extension,
\begin{align}
    \Gamma_{\rm WZ}(g_{\rm cl})
=
-\frac{1}{4\pi}
\int_{T^2}
\tr \theta_m\wedge\theta_w.
\end{align}
On the boundary $r=1$, the condition $f_w(s,1)=e^{isw}$ implies $\left.\theta_w\right|_{r=1}
=
iw\,ds$. Together with $\theta_m=im\,dt$, it gives
\begin{align}
    \Gamma_{\rm WZ}(g_{\rm cl})
=
\pi(m,w).
\end{align}
This is the value for the particular extension \eqref{eq:extension}; changing the extension can shift $\Gamma(g_{\rm cl})$ by $2\pi\mathbb Z$, so the exponentiated amplitude is independent of this choice.
Thus the WZ amplitude for $g_{\rm cl}$ is well defined:
\begin{align}
A_{\rm WZ}(g_{\rm cl})=e^{-i\pi\kappa(m,w)}.
\end{align}

This direct evaluation also admits a natural interpretation in terms of the PW cocycle. Since $g_\sigma$ is a constant left multiplication, it drops out of the Maurer–Cartan form and hence of the WZ functional. Thus
$ \Gamma_{\rm WZ}(g_{\rm cl}) = \Gamma_{\rm WZ}(g_1 g_2)$ with $g_1 = e^{  im t}$ and $g_2 = e^{  i ws}$. Then PW formula~\eqref{eq:PWformula} gives
\begin{align} \label{eq:WZdirectEva}
    \Gamma_{\rm WZ}(g_{\rm cl}) = 
    \Gamma_{\rm WZ}(g_1 ) + \Gamma_{\rm WZ}( g_2 )  + \phi_{\rm WZ}(g_1, g_2).
\end{align}
Since $g_1$ and $g_2$ each depend on only one torus cycle, their WZ functionals vanish. For $g_1=e^{imt}$, choose the filling $D_s^2\times S_t^1$ and extend $g_1$ constantly over $D_s^2$, so the Cartan three-form vanishes. The same argument applies to $g_2=e^{iws}$ using $D_t^2\times S_s^1$. 
Hence
\begin{align}
     \Gamma_{\rm WZ}(g_1 )=\Gamma_{\rm WZ}( g_2 )=0.
\end{align}
and therefore
\begin{align}
    \Gamma_{\rm WZ}(g_{\rm cl})=\phi_{\rm WZ}(e^{  im t}, e^{  i ws})
    =
    \pi(m,w),
\label{eq:GammaWZdirect}
\end{align}
in agreement with the direct evaluation above. Thus the extra phase~\eqref{eq:newphase} is precisely the PW cocycle
\begin{align}
    \xi_{m,w}
    =
    A_{\rm WZ}(g_{\rm cl})
    =
    \Phi_{\rm WZ}(e^{ itm},e^{ isw})
    =
    (-1)^{\kappa(m,w)} .
\end{align}
Thus the phase $\xi_{m,w}$ is the residual global information carried by the WZ gerbe after abelianization: although $\Omega|_T=0$, its PW cocycle on the two winding sectors remains nontrivial.

\subsection{The localization formula}

Having determined the WZ amplitude, we now assemble the remaining
contributions to the localization formula. The two-dimensional terms
in the action evaluated on the localization locus give
\begin{align}
\label{eq:localizationTwodimPart}
\exp\{-\kappa I_{\rm kin}^A(g_{\rm cl})\}
=
H_{m,w}(\tau,\sigma(u),v)
\equiv H^\sigma_{m,w},
\end{align}
where $H_{m,w}$ is given in~\eqref{eq:Hmw}. This agrees with the
corresponding result of Murthy and Witten. Including the WZ amplitude,
the full classical contribution is
\begin{align}
Z_{\rm cl}(g_{\rm cl})
=
\xi_{m,w}\,H^\sigma_{m,w}.
\end{align}

We next consider the one-loop contribution. Its Weyl-group dependence
factorizes as
\begin{align}
Z^{\text{1-loop}}_{\sigma,m,w}
=
\det(\sigma)\,
Z^{\text{1-loop}}_{m,w}.
\end{align}
Indeed, the nonzero-mode fluctuation measure is Weyl invariant, while
the Cartan fermion zero-mode insertion transforms by $\det(\sigma)$. The remaining contribution is
invariant under coroot shifts
$m\to m+\beta$ and $w \to w+\gamma$, with $\beta,\gamma\in Q^\vee$, which
merely relabel the root-space fluctuation modes. Since
$m,w\in Q^\vee$ in the simply connected theory, it follows that $Z^{\text{1-loop}}_{m,w}
=
Z^{\text{1-loop}}_{0,0}$.
Hence
\begin{align}
\label{eq:localizationOneloop}
Z^{\text{1-loop}}_{\sigma,m,w}
=
\frac{\mathrm{Vol}(2\pi\sqrt{\kappa}Q^\vee)}
{(2\pi^2\tau_2)^{r/2}}\,
\det(\sigma)
=
2^r\,\mathrm{Vol}(Q^\vee)
\left(
\frac{\kappa}{2\tau_2}
\right)^{r/2}
\det(\sigma).
\end{align}
Thus the one-loop contribution is independent of the winding data
$m$ and $w$, with all Weyl-group dependence carried by the usual Weyl
sign.

Combining the classical and one-loop contributions, we obtain
\begin{align}
Z_{\rm SWZW}
=
\frac{\mathrm{Vol}(2\pi\sqrt{\kappa}Q^\vee)}
{(2\pi^2\tau_2)^{r/2}}
\sum_{\sigma\in W}\det(\sigma)
\sum_{m,w\in Q^\vee}
\xi_{m,w}\,H^\sigma_{m,w}.
\end{align}
This reproduces the localization formula~\eqref{eq:SWZWPartFunc}.
Thus the localization solutions~\eqref{eq:localizingSol}, the two-dimensional classical contribution~\eqref{eq:localizationTwodimPart}, and the one-loop determinant~\eqref{eq:localizationOneloop} agree with Murthy and Witten~\cite{murthyW25}. The additional factor comes entirely from the WZ amplitude,
\begin{align}
\xi_{m,w}
=
(-1)^{\kappa(m,w)},
\end{align}
and agrees precisely with the phase obtained independently from the
Hamiltonian derivation.

\section{Abelianization and Narain lattice} \label{sec:narain}
The WZW model localizes to a sigma model on the maximal torus by abelianization. Although
the Cartan three-form vanishes upon restriction to the maximal torus, the global
holonomy of the WZ term survives. We will show that this
residual contribution is precisely the Kalb--Ramond (KR) $B$-field coupling of the toroidal
sigma model on the maximal torus. Thus the abelianized WZW theory naturally gives rise to a
Narain CFT at a distinguished point in its moduli space, with the torus
metric and $B$-field determined by the Lie group data.

We first fix some notation. Let $L$ be a rank $d$ lattice in $V\simeq\mathbb R^d$ equipped with a positive-definite inner product $( \cdot,\cdot)$, which identifies $V\simeq V^\ast$. Let $B\in\Lambda^2 V^*$ be a constant 2-form. Choosing a basis $\{e_i\}_{i=1}^d$ of $L$, we write $ G_{ij}=(e_i,e_j)$, $B_{ij}=B(e_i,e_j)$, 
so that $G_{ij}=G_{ji}$ and $B_{ij}=-B_{ji}$. 
Let ${f^i}$ be the basis of $V$ dual to ${e_i}$, $(f^i,e_j)=\delta^i_j$. Regarded as a linear map $B:V\to V$ via $V^*\simeq V$, we then have $Be_j=\sum_i B_{ij}f^i$, by linearity, this defines $B\gamma\in V$ for every $\gamma\in L$.

\subsection{Abelian SWZW model and Siegel--Narain theta function}

We first consider the abelian WZW model for $G = U(1)^d$, namely the Narain sigma model with torus target $T^d=V/(2\pi L)$ with metric $G_{ij}$ and a constant Kalb--Ramond (KR) $B$-field $B_{ij}$. The fields are $X=X^i e_i$ with periodicity $ X^i \sim X^i +  2 \pi n^i$, $n^i\in\mathbb Z$.
 The action of Narain--Sarmadi--Witten (NSW) is~\cite{Narain:1986am}
\begin{align}
    S_{\rm NSW}(X)
    =
    \frac{1}{8\pi}
    \left(
    \int_{T^2}
    G_{ij}\,dX^i\wedge *dX^j
    +
    i\int_{T^2}
    B_{ij}\,dX^i\wedge dX^j
    \right)
\end{align}
or, equivalently in complex coordinate
\begin{align}
    S_{\rm NSW}(X)
    =
    \frac{i}{4\pi}
    \int_{T^2} dz\wedge d\bar z\,
    (G_{ij}+B_{ij})\,
    \partial X^i\bar\partial X^j .
\end{align}
The global $B$-field holonomy, which is important for the identification with the abelianized WZW model, is described in Appendix~\ref{apx:WZ&Bfield}.

The action has a $U(1)^d_L\times U(1)^d_R$ symmetry acting on $X$ by
\begin{align}
    \delta X=\epsilon_L-\epsilon_R .
\end{align}
We couple this symmetry to background gauge fields $A_L$ and $A_R$, with covariant derivative
\begin{align}
    DX=dX+A_L-A_R .
\end{align}
Then the gauged Narain action is given by
\begin{align}
S_{\rm NSW}(X,A_L,A_R)
=
\frac{1}{8\pi}\int_{T^2}
G_{ij}DX^i\wedge *DX^j
+
\frac{i}{4\pi}\int_{T^2} X^{\ast} (B_{\rm eq}).
\end{align}
or equivalently in complex coordinate
\begin{align}
S_{\rm NSW}(X,A_L,A_R)
& =
\frac{i}{4\pi}
\int_{T^2} dz\wedge d\bar z
(G_{ij}
+
B_{ij})\partial X^i\bar\partial X^j\\
& \qquad + \frac{i}{2 \pi}
\int_{T^2} dz\wedge d\bar z
G_{ij}\big( A_{z}^{L,i} \bar\partial X^j
-
\partial X^j A^{R, i}_{\bar z}
-
A^{L,i}_z A^{R,j}_{\bar z} \\
& \hspace{5cm} + \frac{1}{2 } (A^{L,i}_z A^{L,j}_{\bar z} + A^{R,i}_z A^{R,j}_{\bar z})
\big ).
\end{align}
The corresponding gauged $B$-field coupling in terms of equivariant extension $B_{\rm eq}$ is discussed in Appendix~\ref{apx:WZKRgauging}.

We now supersymmetrize the model. Since the $U(1)^d_L\times U(1)^d_R$ symmetry acts by shifts of $X$, the fermions are neutral, and their covariant derivatives reduce to ordinary derivatives.  For flat background gauge fields~\eqref{eq:ALAR}, the abelian SWZW action is
\begin{align}
S_{\rm SNSW}(X, \psi, \tilde \psi; A^L,A^R)
 =
S_{\rm NSW}(X; A^L,A^R) + S_F(\psi) + S_F(\tilde \psi), 
\end{align}
with
\begin{align}
   S_F(\psi) =  -
\frac{1}{4\pi}
\int_{T^2}dz \wedge d\bar z\,
G_{ij}
\psi^i \partial_{\bar{z}}\psi^j  ,  \quad 
   S_F(\tilde \psi)= -
\frac{1}{4\pi}
\int_{T^2}dz \wedge d\bar z\,
G_{ij}
\widetilde\psi^i \partial_z \widetilde\psi^j.
\end{align}
The action is invariant under the supersymmetries
\begin{align}
    QX^i = \psi^i, \quad Q\psi^i=iD_z X^i, \quad Q \tilde \psi^i= 0,  \quad  QA^L= QA^R =0 
\end{align}
and
\begin{align}
    \tilde QX^i = \tilde \psi^i, \quad \tilde Q\psi^i= 0, \quad \tilde Q \tilde \psi^i= iD_{\bar z} X^i,  \quad \tilde Q A^L= \tilde Q A^R =0 
\end{align}
which satisfies $Q^2 = iD_z$ and $\tilde Q^2 = iD_{\bar z}$.

We impose periodic (Ramond–Ramond) boundary conditions on the fermions and saturate their constant zero modes $\psi_0^i$ and $\tilde\psi_0^i$ by the insertion
\begin{align}
Z_{\rm SNSW}
=
\int\mathcal{D}X \mathcal{D}\psi \mathcal{D}\tilde\psi
(\psi_0^1\cdots\psi_0^d)
(\tilde\psi_0^1\cdots\tilde\psi_0^d)
e^{-S_{\rm SNSW}}.
\end{align}
For localization, we deform the action by a $Q$-exact term,
\begin{align}
S_{\rm SNSW}\rightarrow S_{\rm SNSW}+\lambda\,Q V ,
\end{align}
with the localization term
\begin{align}
V
=
 \int dz \wedge d \bar z\,
G_{ij}\,
\partial_{\bar z} \psi^i
\,\pd_z \partial_{\bar z}X^j .
\end{align}
such that
\begin{align}
Q V
=
i \int dz \wedge d \bar z\,
G_{ij}
\partial_{\bar z} \partial_{z}X^i
\,\pd_z \partial_{\bar z}X^j
- \int dz \wedge d \bar z\,
G_{ij}
\partial_{\bar z} \psi^i
\,\pd_z \partial_{\bar z}\psi^j .
\end{align}
Taking $\lambda\to\infty$ gives the localization equation
\begin{align}
\partial_{\bar z}\partial_z X^i=0,
\end{align}
which is precisely the classical equation of motion in the presence of flat background gauge fields. Thus, on the torus, the localization locus consists of the classical solutions
\begin{align}
X^i_{\rm cl}=m^i t+w^i s,
\qquad
m,w\in L,
\end{align}
together with $\psi_{\rm cl}=0$.

Expanding around the localization locus as
\begin{align}
    X=X_{\rm cl}+\frac{Y}{\sqrt{\lambda}},
\quad
\psi\rightarrow\frac{\psi}{\sqrt{\lambda}},
\end{align}
while leaving $\tilde\psi$ unscaled, gives
\begin{align}
    S_{\rm SNSW}+\lambda QV
=
S_{\rm NSW}(X_{\rm cl};A^L,A^R)
+S_F(\tilde\psi)
+(QV)^{(2)}
+O(\lambda^{-1/2}).
\end{align}
The partition function therefore localizes to
\begin{align}
Z_{\rm SNSW}
=
\sum_{m,w\in L}
Z^{\rm cl}_{m,w}
Z^{\rm 1-loop},
\end{align}
with
\begin{align}
Z^{\rm cl}_{m,w}
=
e^{-S_{\rm NSW}(X_{\rm cl},A^L,A^R)},\quad 
Z^{\rm 1-loop}
=
\int\mathcal D\tilde\psi
e^{- S_F(\tilde\psi)}
\int\mathcal DY\mathcal D\psi
e^{-(QV)^{(2)}}.
\end{align}
Notice that the one-loop contribution is independent of $m$ and $w$. 
The nonzero bosonic and fermionic fluctuations cancel pairwise, leaving only the zero-mode contribution. Therefore the one-loop contribution is
\begin{align}
    Z^{\text{1-loop}}
    =
    \frac{{\rm Vol}(2\pi L)}
    {(2\pi^2\tau_2)^{d/2}},
\end{align}
where $\text{Vol}(L) = \det(G)^{1/2}$.
The classical contribution evaluates to
\begin{align}
Z^{\rm cl}_{m, w}
=
\exp\{-
\pi i B(m, w) -\frac{\pi}{2 \tau_2} \big( 
(m-\tau w+2u, m-\bar\tau w-2 \bar v ) \notag \\ 
+ 2 (u,\bar v) + (u, \bar u) + (\bar v, v) \big)  \}.
\end{align}
Therefore the localization formula of the partition function is 
\begin{align}
Z_{\rm SNSW}
& =\frac{{ \rm Vol}(2 \pi L)}{(2\pi^2\tau_2)^{d/2}}
\sum_{m,w\in L}
\exp\{-
\pi i B(m, w)  \notag \\ 
&\qquad \qquad  -\frac{\pi}{2 \tau_2} \big( 
(m-\tau w+2u, m-\bar\tau w-2 \bar v ) 
+ 2 (u,\bar v) + (u, \bar u) + (\bar v, v) \big)  \}.
\end{align}

The localization result can be naturally reorganized in terms of the left- and right-moving charge spectrum of the toroidal CFT. The $U(1)^d_L\times U(1)^d_R$ symmetry enhances to the affine current algebra
\begin{align}
     \widehat{\mathfrak u}(1)^d_L\times
 \widehat{\mathfrak u}(1)^d_R.
\end{align}
Decomposing the compact bosons into their left- and right-moving components,
\begin{align}
X^i(z,\bar z)=X_L^i(z)+X_R^i(\bar z),
\end{align}
the corresponding affine currents are
\begin{align}
    J_L^i(z)=i\partial X^i_L(z),
    \quad
    J_R^i(\bar z)=i\bar\partial X^i_R(\bar z).
\end{align}
The primary vertex operators are
\begin{align}
    V_{p_L,p_R}
    =
    :\exp\!\left[
    i(p_L,X_L)+i(p_R,X_R)
    \right]: ,
\end{align}
where the left- and right-moving momenta are labeled by
$(\lambda,\gamma)\in L^\ast\oplus L$,
\begin{align}
    p_L
    =
    \lambda+\frac12(B\gamma -\gamma),
    \qquad
    p_R
    =
    \lambda+\frac12(B\gamma +\gamma).
\end{align}
The momentum spectrum therefore forms the Narain lattice
$\Gamma \subset V\oplus V$.
Equipped with the Lorentzian bilinear form
\begin{align}
    \langle p,p'\rangle
    =
    (p_L,p_L')-(p_R,p_R'),
\end{align}
one finds
\begin{align}
    \langle p,p\rangle
    =
    p_L^2-p_R^2
    =
   - 2(\lambda,\gamma)\in2\mathbb Z .
\end{align}
Thus $\Gamma$ is even; moreover, the duality between $L^{\ast}$ and $L$ implies that $\Gamma$ is
self-dual.

We can now see directly how this charge lattice emerges from the
localization sum. Poisson resummation over the momentum variable
$m\in L$ converts the winding representation into a
momentum--winding representation,
\begin{align}
\label{eq:Zsnsw}
Z_{\rm SNSW}(\tau,u,v)
&=
2^d\,C
\sum_{p\in\Gamma}
q^{p_L^2/2}\,
\bar q^{p_R^2/2}\,
e^{2 \pi i (p_L,u)} e^{2\pi i (p_R, -\bar v)},
\end{align}
where $\Gamma$ is the Narain lattice for the charge spectrum determined
above, and
\begin{align}
 C
=
\exp\{
\frac{\pi i}{\tau_2}
\left(
(u ,{\rm Im} u)
+ 
(\bar v,{\rm Im} \bar v)
\right)
\}.
\end{align}
This motivates the definition of the Siegel--Narain theta function
associated with $\Gamma$,
\begin{align}
\mathcal S_\Gamma
(\tau,\bar\tau, \theta)
&=
e^{E(\tau, \bar \tau, \theta )}
\sum_{p\in\Gamma}
 e^{\pi i \tau \langle p, p \rangle_{+}  + \pi i \bar \tau \langle p, p \rangle_{-} +2 \pi i \langle p, \theta \rangle }
\label{eq:SN-theta}
\end{align}
where $\theta=[\theta_L;\theta_R]\in\Gamma\otimes\mathbb C$, and 
\begin{align}
E(\tau, \bar \tau, \theta )
=
\frac{\pi i}{\tau_2}
\left(
\langle \theta , {\rm Im} \theta \rangle_{+}
-
\langle \theta , {\rm Im} \theta \rangle_{-}
\right).
\label{eq:SN-completion}
\end{align}
Thus
\begin{align}
Z_{\rm SNSW}(\tau,u,v)
=
2^d\, \mathcal S_\Gamma(\tau,\bar\tau,[u;\bar v]).
\end{align}
For the even self-dual Narain lattice $\Gamma$, the Siegel--Narain
theta function transforms as a modular form of weight $(d/2,d/2)$.

\subsection{Abelianization, $B$-field holonomy, and the Narain description} 

Let us now compare the classical localization contribution of the SWZW model with that of the gauged Narain sigma model. The localization equations are imposed on the full WZW field $g$, but every localization solution can be represented by a harmonic Cartan-valued field. In the sector labeled with $m,w \in Q^\vee$ we write
\begin{align}
  g_{\rm cl} =  g_{\sigma} e^{iX_{\rm cl}} \, ,\quad   X_{\rm cl} =  mt + ws.
\end{align}
On the localization locus, the kinetic and gauging terms of the WZW action agree with those of the NSW model with
$L= Q^\vee, G= \kappa \mathsf C$. The only remaining possible discrepancy is therefore the topological term. Comparing the WZ amplitude with the KR amplitude gives
\begin{align}
\frac{A_{\rm WZ}(g_{\rm cl})}{A_{\rm KR}(X_{\rm cl})} = \exp\{ - \pi i G_{ij} m^i w^j + \pi i\,B_{ij}m^i w^j\}
\end{align}
Thus the two amplitudes agree provided $B\equiv G\pmod 2$. A representative is given by $B  = \kappa \mathsf{B}$ with
\begin{align}
    \mathsf{B}_{ij} = \begin{cases}
      \mathsf{C}_{ij}  & i < j\\
        - \mathsf{C}_{ij}  & i>j
    \end{cases}
\label{eq:Brepresentative}
\end{align}
with $\mathsf C_{ij}$ defined in~\eqref{eq:GramMatrix}.
Equivalently, decomposing $\mathsf{C} =  D + U + U^{T}$ where $D$ is diagonal and $U$ is strictly upper triangular, then we have $\mathsf{B}  = (U-U^T)$. Consequently, the abelianized WZW model corresponds not to a generic point of the Narain moduli space, but to the distinguished point determined by $(Q^\vee,\kappa \mathsf C, \kappa \mathsf B)$. 

This identifies the global WZ holonomy with the flat $B$-field holonomy of the torus theory. Although $\Omega|_T=0$, the global holonomy need not vanish; it survives precisely as the lattice phase produced by the $B$-field. In this sense, abelianization converts the topological WZ term into the $B$-field data of the Narain model.

We can now identify the charge lattice appearing in the localized partition function by comparing~\eqref{eq:McZexpanded} with the Siegel–Narain theta function~\eqref{eq:SN-theta}. With
\begin{align}
L=Q^\vee,
\qquad
L^*=\frac{1}{\kappa}P,
\end{align}
the comparison identifies
\begin{align}
p_L = \frac{\lambda}{\kappa},
\qquad
p_R = \frac{\lambda}{\kappa}+w,
\qquad
\lambda\in P,\quad
w\in Q^\vee.
\end{align}
leading to the charge lattice
\begin{align}
\Gamma = \left\{ p=[p_L;p_R]\in L^*\times L^* : p_R-p_L\in L \right\}.
\end{align}
With respect to the bilinear form induced by $G=\kappa\mathsf C$, $\Gamma$ is an even self-dual lattice of signature $(r,r)$, and
\begin{align}
\mathcal Z(\tau,u,v) = 2^r \, \mathcal S_\Gamma(\tau,\bar\tau,[u;\bar v]).
\end{align}
The $B$-field gives the standard Narain reparametrization of these momenta. In the present case,
\begin{align}
\frac{1}{2}(1+\mathsf B)Q^\vee\subset P,
\end{align}
so that the shift
\begin{align}
\lambda
\mapsto
\lambda+\frac{\kappa}{2}(-1+\mathsf B)w
\end{align}
preserves the weight lattice $P$. The momenta may therefore be written as
\begin{align}
p_L
=
\frac{\lambda}{\kappa}
+\frac{1}{2}(-1+\mathsf B)w,\quad 
p_R
=
\frac{\lambda}{\kappa}
+\frac{1}{2}(1+\mathsf B)w,
\quad
\lambda\in P,\quad
w\in Q^\vee.
\end{align}
This is precisely the Narain momentum spectrum associated with the distinguished Narain data
$(Q^\vee,\kappa\mathsf C,\kappa\mathsf B)$.

The remaining step is to recover the full SWZW theory from this Narain description, by summing over the Weyl group images of the Cartan localization locus. Hence~\eqref{eq:ZswzwinMcZ} becomes
\begin{align}
Z_{\rm SWZW}
=
2^r
\sum_{\sigma\in W}
\det(\sigma) \, \mathcal S^\sigma_\Gamma,
\quad
\mathcal S^\sigma_\Gamma(\tau,u,v)
=
\mathcal S_\Gamma(\tau,\sigma(u),v).
\end{align}
Equivalently,
\begin{align}
Z_{\rm SWZW} = \sum_{\sigma\in W} \det(\sigma) \, Z_{\rm SNSW}^{\sigma}.
\end{align}

For the bosonic WZW model, the corresponding decomposition takes the form
\begin{align}
Z_{\rm WZW} = \frac{Z_{\rm SWZW}}{\rm Z_F}
=
\sum_{\sigma\in W}
Z_{\rm NSW}^{\sigma}\,
Z_{\beta\gamma,\mathfrak g/\mathfrak h}^{\sigma},
\end{align}
where 
\begin{align}
Z_{\rm NSW}
=
\frac{\mathcal S_\Gamma}{|\eta(\tau)|^{2r}},
\end{align}
and the non-Cartan one-loop determinant is
\begin{align}
Z_{\beta\gamma,\mathfrak g/\mathfrak h}
=
\,
{\rm Pf}^{-1}_{\mathfrak g/\mathfrak h}
(\partial_z+A_z^R)\,
{\rm Pf}^{-1}_{\mathfrak g/\mathfrak h}
(\partial_{\bar z}+A_{\bar z}^L).
\end{align}
Under a Weyl transformation,
\begin{align}
Z_{\beta\gamma,\mathfrak g/\mathfrak h}^{\sigma}
&=
{\rm Pf}^{-1}_{\mathfrak g/\mathfrak h}
(\partial_z+A_z^R)\,
{\rm Pf}^{-1}_{\mathfrak g/\mathfrak h}
(\partial_{\bar z}+g_\sigma^{-1}A_{\bar z}^L g_\sigma)
\nonumber\\
&=
\det(\sigma)\,
Z_{\beta\gamma,\mathfrak g/\mathfrak h}.
\end{align}
The factor $Z_{\beta\gamma,\mathfrak g/\mathfrak h}$ is the one-loop
determinant of the fluctuations transverse to the maximal torus, namely
along $G/T$. Integrating out these non-Cartan modes leaves an effective
theory entirely in the Cartan subalgebra, while their determinant supplies
the Weyl sign $\det(\sigma)$. It would be interesting to relate this
localization picture more directly to abelianization by singular gauge
transformations~\cite{blauT93,Blau:1994rk,blauT93lecture}, where
diagonalization similarly reduces the theory to the maximal torus with a
residual Weyl sum.

\section{Particle limit and Frenkel's formula} \label{sec:frenkel}
We expect the quantum mechanics of a particle moving on the group manifold $G$ to arise from the WZW model by shrinking the spatial circle. In this section, we implement this dimensional-reduction limit directly in the torus partition function and show that the resulting expression reproduces Frenkel’s trace formula. 

We consider a torus with spatial circumference $L$ and Euclidean time circumference $\beta$, and take
\begin{align} \label{eq:particleLimit}
    \tau=i\frac{\beta}{L},\qquad
    k=\frac{2\pi}{L},
\end{align}
and the \emph{particle limit} is obtained by sending $L\to 0$. In this limit the level $k$ goes to infinity, while the Euclidean time circumference remains fixed. This is the \emph{large-level}, or semiclassical, limit in which spatially varying configurations become infinitely costly, and the dynamics reduces to quantum mechanics on the group manifold $G$. Equivalently, the path integral is dominated by configurations that are constant along the spatial circle and vary only along Euclidean time. The resulting quantum-mechanical spectrum consists of $(\lambda, \lambda)$, where $\lambda$ runs over all dominant weights of $G$~\cite{gepnerWitten86}.

It is worth noting that the localization and Hamiltonian descriptions exhibit the same underlying mechanism. In both descriptions, nontrivial lattice sectors contribute terms whose real exponent contains a negative contribution of order $L^{-2}$, and are therefore exponentially suppressed in the particle limit. In the localization description, these sectors are labeled by nonzero winding, whereas in the Hamiltonian description they correspond to nonzero lattice shifts in the affine theta functions. Thus, in both descriptions, the particle limit projects onto the sectors that remain finite under dimensional reduction. The same argument applies for complex $\beta$ with ${\rm Re}(\beta)>0$, where the nontrivial lattice sectors remain exponentially suppressed.

\subsection{Particle limit in the Hamiltonian formulation}

In the Hamiltonian description, the SWZW partition function is written as
\begin{align} 
    Z_{\text{SWZW}}(\tau, u, v) = 2^r C_{\kappa} \sum_{\lambda \in P^k_{+}} N_{\lambda + \rho,\kappa}(\tau, u )  N_{\lambda + \rho,\kappa}(- \bar{\tau}, - \bar v)  . 
\end{align}
The prefactor $C_{\kappa}$ has the following finite limit:
\begin{align}
\mathcal{C}_{\infty} = \lim_{L\to0}C_\kappa
=
\exp\{ 
\frac{2\pi^2 i}{\beta}
\left(
(u, {\rm Im}u)
+
(\bar v, {\rm Im}\bar v)
\right)
\}.
\end{align}
We next consider the Weyl--Kac numerator
\begin{align}
N_{\lambda + \rho,\kappa}
 =   \sum_{\sigma \in W} \det(\sigma) \Theta_{\sigma(\lambda + \rho), \kappa}
  =   \sum_{\sigma \in W} \det(\sigma) \sum_{\gamma \in Q^\vee} f_{\sigma,\gamma}, 
\end{align}
where
\begin{align}
    f_{\sigma,\gamma} = \exp\{\frac{\pi i \tau}{\kappa} (\sigma(\lambda + \rho) + \kappa \gamma)^2+2\pi i\left(\sigma(\lambda + \rho) + \kappa \gamma,u\right)\}.
\end{align}
Since $\kappa\sim L^{-1}$, the affine lattice translate $\kappa\gamma$ scales as $L^{-1}\gamma$. For $\gamma\neq0$, the quadratic term in the theta-function exponent therefore produces a negative contribution of order $L^{-2}$ to its real part. These sectors are exponentially suppressed, whereas the $\gamma=0$ sector has a finite limit. Consequently,
\begin{align}
    \lim_{L\to0} f_{\sigma,\gamma} = \delta_{\gamma,0}
\exp\{
-\frac{\beta}{2}(\lambda+\rho, \lambda+\rho)
+
2\pi i\bigl(\sigma(\lambda+\rho),u\bigr)
\}.
\end{align}
Hence, the Weyl--Kac numerator admits the limit
\begin{align}
    \lim_{L \to 0} N_{\lambda + \rho, \kappa}   = e^{-\frac{1}{2} \beta (\lambda+\rho, \lambda+\rho)}  \sum_{\sigma \in W} \det(\sigma) e^{2\pi i \left(\sigma(\lambda+\rho),u\right)} =  e^{-\frac{1}{2} \beta  (\lambda+\rho, \lambda+\rho)} J_{\lambda + \rho} .
\end{align}
For every fixed $\lambda\in P_+$, the weight $ \lambda$ belongs to $P_+^k$ for all sufficiently large $k$. Thus in the large-level limit, the sum over $\lambda \in P_+^k$ becomes a sum over all dominant weights $\lambda \in P_+$. Hence the partition function admits the limit
\begin{align} 
\label{eq:QMchractersum}
\lim_{L\to0}Z_{\mathrm{SWZW}}
 =
2^r \mathcal{C}_{\infty} 
\sum_{\lambda\in P_+}
e^{-\beta(\lambda+\rho,\lambda+\rho)}
J_{\lambda+\rho}(u)
J_{\lambda+\rho}(-\bar {v}). \notag
\end{align}
Notice that the factor $2^r$, which arises from the quantization of the neutral Majorana zero modes, is carried unchanged through the particle limit.

\subsection{Particle limit for the localization formula}

We now take the same particle limit in the localization formula for the SWZW partition function,
\begin{align}
   Z_{\text{SWZW}}=  2^r
{\rm Vol}(Q^\vee)
\left(
\frac{\kappa}{2\tau_2}
\right)^{\frac{r}{2}}\sum_{\sigma\in W} \det(\sigma ) \sum_{m,w\in Q^\vee}  (-1)^{\kappa (m, w)} H^{\sigma}_{m,w},
\end{align}
where
\begin{align}
    H^{\sigma}_{m,w}  =   \exp \{ -\frac{\pi \kappa}{2 \tau_2} \big ( 
                 (m- \tau w  + 2 \sigma(u),\, m - \overline{\tau} w  - 2 \bar{v}) 
                  + \, 2 (\sigma(u),\bar{v}) + (u, \bar{u}) + (\bar{v},v)
              \big ) \}.
\end{align}
For $w\neq0$, the quadratic term in $\tau w$ and $\bar\tau w$ produces a negative contribution of order $L^{-2}$ to the real part of the exponent. Hence all nonzero winding sectors are exponentially suppressed, while the $w=0$ sector remains finite. Therefore,
\begin{align} 
    \lim_{L\to0} H^{\sigma}_{m,w}
    =
    \delta_{w,0} H^{\sigma}_{m,0},
\end{align}
where 
\begin{align}
    H^{\sigma}_{m,0} 
  &  =   \exp \{ -\frac{\pi^2}{\beta} 
    \big ( 
     (m   + 2 \sigma(u),\, m   - 2 \bar{v})   
     + \, 2 (\sigma(u),\bar{v}) + (u, \bar{u}) + (\bar{v},v)
    \big ) \} \\
 & = \mathcal{C}_{\infty} e^{-\frac{\pi^2}{\beta} \left( m+ \sigma(u) -\bar  v \right)^2
}.
\end{align}
For the surviving $w=0$ sector, the phase factor is simply $(-1)^{\kappa(m,0)}=1$. 
Thus the localization formula reduces in the particle limit to
\begin{align} \label{eq:QMlocalization}
\lim_{L\to0}Z_{\mathrm{SWZW}}
  & = 2^r \mathcal{C}_{\infty }\left(\frac{\pi}{\beta}\right)^{r/2}
{\rm Vol}(Q^\vee) \sum_{\sigma\in W}\det(\sigma) \sum_{m\in Q^\vee}
e^{
-\frac{\pi^2}{\beta}
\left(
m+ \sigma(u) -\bar v
\right)^2
}.
\end{align}

This can be compared with the localization formula of Choi and Takhtajan for supersymmetric quantum mechanics on the group manifold $G$~\cite{choiT25}. Their approach considers $\mathcal N=1$ supersymmetric quantum mechanics, with a single adjoint Majorana fermion charged under $G_R$. Hence it could be obtained by dimensional reduction of the two-dimensional $\mathcal N=(0,1)$ WZW model.

\subsection{Frenkel's heat kernel trace formula}

We can now compare the two forms of the particle-limit partition function and obtain the equivalent form of Frenkel’s formula
\begin{align}
\sum_{\lambda\in P_{++}}
J_{\lambda}(u)
J_{\lambda}(-\bar v) \,
e^{-\beta \lambda^2}
&=
\left(\frac{\pi}{\beta}\right)^{r/2}
{\rm Vol}(Q^\vee)
\sum_{\sigma\in W}
\det(\sigma)
\sum_{m\in Q^\vee}
e^{
-\frac{\pi^2}{\beta}
\left(
m+\sigma(u)-\bar v
\right)^2
}.
\label{eq:FrenkelSUSY}
\end{align}
Using the Weyl character formula
\begin{align}
\chi_\lambda(u)
&=
\frac{J_{\lambda+\rho}(u)}{J_\rho(u)},
\end{align}
and
\begin{align}
c_2(\lambda)
&=
(\lambda+\rho)^2-\rho^2,
\end{align}
this becomes the familiar Frenkel formula for the heat-kernel~\cite{Frenkel1984}
\begin{align} 
\label{eq:Frenkel}
\sum_{\lambda\in P_+}
\chi_\lambda(u)
\chi_\lambda(- \bar v) e^{-\beta c_2(\lambda)}
& = 
\frac{
\left(\frac{\pi}{\beta}\right)^{r/2}
{\rm Vol}(Q^\vee)
e^{\beta(\rho,\rho)}
}{
J_\rho(u)
J_\rho(-\bar v)
}
\sum_{\sigma\in W}
\det(\sigma)
\sum_{m\in Q^\vee}
e^{
-\frac{\pi^2}{\beta}
\left(
m + \sigma(u) -\bar  v
\right)^2
}. 
\end{align}
Both sides define holomorphic functions for $\beta\in\mathbb{C}$ with ${\rm Re}(\beta)>0$, and $u,v\in\mathfrak{h}_{\mathbb{C}}$.
It equates two descriptions of the same quantum mechanics on $G$. In the Hamiltonian description, the Hilbert space is given by the Peter–Weyl decomposition
\begin{align}
L^2(G)
\simeq
\bigoplus_{\lambda\in P_+}
V_\lambda\otimes V_\lambda^*.
\end{align}
With our convention for the Laplace--Beltrami operator,
\begin{align}
    \Delta_G\chi_\lambda=c_2(\lambda)\chi_\lambda,
\end{align}
the Euclidean evolution operator $e^{-\beta\Delta_G}$ contributes the factor $e^{-\beta c_2(\lambda)}$ in the representation $V_\lambda\otimes V_\lambda^\ast$.
In the localization description, the same quantity is represented by a Weyl-antisymmetrized sum over the coroot lattice. The equality between these two descriptions—the sum over irreducible representations and the affine Weyl group sum--is precisely Frenkel’s formula.

It remains to verify directly that the particle limit of the WZW affine character expansion reproduces the left-hand side of~\eqref{eq:Frenkel}. This requires some care with the normalization of affine characters. Three closely related conventions will be useful.
\begin{align}
   \mathcal{U}_{K,k} & = {\rm Tr}_{V_{K,k}}(q^{L_0})  = \frac{\mathcal{N}_{K+\rho, k+h^\vee}}{\mathcal{N}_{\rho,h^\vee}}  ,\\
   \chi_{K,k} & =  {\rm Tr}_{V_{K,k}}(q^{L_0-\frac{c_{\mathfrak{g},k}}{24}}) =\frac{N_{K + \rho ,k+h^\vee}}{N_{\rho,h^\vee}},\\
    {\rm ch}_{\hat K} & =  {\rm Tr}_{V_{K,k}}(q^{L_0-\frac{c_2(K)}{2(k +h^{\vee})}})  =\frac{J_{\hat K + \hat \rho}}{J_{\hat\rho}} . 
\end{align}
where
\begin{align}
   c_{\mathfrak{g},k}
   =\frac{k \dim(\mathfrak g)}{(k+h^\vee)},
\end{align}
and
\begin{align}
    J_{\hat{\lambda}+\hat{\rho} }
= \sum_{w \in \hat{W}} \det(w) w( e^{\hat{\lambda}+\hat{\rho}  }) ,
\end{align}
where $\hat{\rho} = \rho + h^\vee \Lambda_0$ and $\hat{W}$ is the affine Weyl group, and 
\begin{align}
\label{eq:WeylInvKacNum}
\mathcal{N}_{K+\rho, k+h^\vee}(\tau, v) =  \frac{q^{-\frac{(\rho,\rho)}{2(k +h^{\vee})}} }{J_\rho(v)}N_{K+\rho,\,k+h^\vee }(\tau , v),
\end{align}
These are related by
\begin{align}
    \mathcal{U}_{K,k} = q^{\frac{c_{\mathfrak{g},k}}{24}} \chi_{K,k} = q^{\frac{c_2(K)}{2(k +h^{\vee})}} {\rm ch}_{\hat{K}} , \quad \chi_{K,k} = q^{h_K} {\rm ch}_{\hat{K}}.
\end{align}
where
\begin{align}
       h_K = \frac{(K + \rho)^2}{2(k +h^{\vee})} - \frac{\rho^2}{2h^{\vee}} =  \frac{c_2(K)}{2(k +h^{\vee})} - \frac{c_{\mathfrak{g},k}}{24}, 
\end{align} 
These affine characters have the expansions
\begin{align}
    \mathcal{U}_{K,k}& = q^{ \frac{c_2(K)}{2(k +h^{\vee})}}  ( \chi_K(u) + O(q)), \\
    \chi_{K,k} & =q^{h_K} ( \chi_K(u) + O(q)),\\
    {\rm ch}_{\hat{K}} & = \chi_K(u) + O(q)  .
\end{align}

The factors $q^{c_{\mathfrak g,k}/24}$ and $q^{h_K}$ separately have singular particle limits, reflecting the singular behavior of $q^{-\frac{(\rho,\rho)}{2h^\vee}}$. Their combination appearing in $\mathcal U_{K,k}$, however, has a finite limit:
\begin{align}
  \lim_{L\to 0}  q^{ \frac{c_2(K)}{2(k +h^{\vee})}} =
e^{-\frac{\beta c_2(K)}{2}}.
\end{align}
Therefore
\begin{align}
  &   (q\bar q)^{\frac{c_{\mathfrak{g},k}}{24}} \sum_{\lambda\in P^k_{+}} \chi_{\lambda,k} \,\bar{\chi}_{\lambda,k} = \sum_{\lambda\in P^k_{+}} \mathcal{U}_{\lambda,k} \,\bar{\mathcal{U}}_{\lambda,k} = \sum_{\lambda\in P^k_{+}}  (q\bar q)^{ \frac{c_2(\lambda)}{2(k +h^{\vee})}} {\rm ch}_{\lambda,k} \,\bar{\rm ch}_{\lambda,k} \\
   \overset{L\to 0}{\longrightarrow}\quad & \sum_{\lambda\in P_+}
\chi_\lambda(u)
\chi_\lambda(- \bar v) e^{-\beta c_2(\lambda)}.
\end{align}
This is precisely the Peter--Weyl character expansion of the heat kernel
on $G$. Thus the particle limit of the WZW partition function reproduces
the Hamiltonian side of Frenkel's formula.

\subsection{Kac--Weyl numerator formula and heat equation}

We now use Frenkel’s formula~\eqref{eq:FrenkelSUSY} to make explicit the relation between the Kac--Weyl numerator and the heat kernel on $G$.  The mixed Virasoro $\times$ Kac–Moody Ward identities imply that the Kac--Weyl numerator satisfies a heat equation~\cite{Bernard:1987df}. This representation will also make the particle limit of the WZW partition function transparent. To this end, we specialize
\begin{align}
    \beta =-\frac{\pi  i \tau }{\kappa }, \quad u=\frac{\Lambda }{\kappa }, \quad  \kappa \in \mathbb{Z}_{+} , \quad\Lambda \in P^{\kappa }_{++}
\end{align}
and replace $-\bar v$ with $v$.  
Under this specialization, the lattice side of Frenkel’s formula can be identified with the modular $S$-transform of the Weyl–Kac numerator~\eqref{eq:WeylKacNum}. In particular,
\begin{align}
&\left(-\frac{\kappa}{i\tau}\right)^{r/2}
{\rm Vol}(Q^\vee)
\sum_{\sigma\in W}
\det(\sigma)
\sum_{m\in Q^\vee}
\exp\left\{
-\frac{\pi i\kappa}{\tau}
\left(m+\sigma(u)+v\right)^2
\right\}
\nonumber\\
= &\,
\left(-\frac{\kappa}{i\tau}\right)^{r/2}
{\rm Vol}(Q^\vee) \,
e^{-\frac{\pi i\kappa v^2}{\tau}} \,
N_{\Lambda,\kappa}
\left(-\frac{1}{\tau},-\frac{v}{\tau}\right)
=
\sum_{L\in P_{++}^{\kappa}}
J_L\left(\frac{\Lambda}{\kappa}\right)
N_{L,\kappa}(\tau,v),
\end{align}
where we used the modular transformation
\begin{align}
\label{eq:Stranform}
N_{\Lambda,\kappa}
\left(-\frac1\tau,-\frac{v}{\tau}\right)
=
(-i\tau)^{r/2}
e^{\pi i\kappa v^2/\tau}
\sum_L S_{\Lambda L}N_{L,\kappa}(\tau,v),
\end{align}
with
\begin{align}
S_{\Lambda L}
= \kappa^{-r/2}{\rm Vol}(Q^\vee)^{-1}J_L\left(\frac{\Lambda}{\kappa}\right)
.
\end{align}

On the other hand, on the representation-theoretic side, we have
\begin{align}
\sum_{\lambda \in P_{++}}
J_\lambda( \frac{\Lambda}{\kappa})
J_\lambda( v)
q^{\frac{\lambda^2}{2\kappa}}
=
\sum_{\lambda \in P_{++}}
\det(\sigma_\lambda)\,
J_{\hat\sigma_\lambda(\lambda)}
(\frac{\Lambda}{\kappa})
J_\lambda(v)\,
q^{\frac{\lambda^2}{2\kappa}} .
\end{align}
where for each $\lambda \in P_{++}$, we choose the unique affine Weyl element $\hat\sigma_\lambda=(\sigma_\lambda,\kappa\gamma_\lambda)$ such that $\hat\sigma_\lambda(\lambda) = \sigma_\lambda(\lambda) + \kappa\gamma_\lambda \in P_{++}^{\kappa}$. Here we have used 
\begin{align}
    J_{\sigma(\lambda)} = \det(\sigma) J_\lambda,\quad J_{\lambda + \kappa\gamma}(\frac{\Lambda}{\kappa}) = J_{\lambda}(\frac{\Lambda}{\kappa}).
\end{align}

Comparing this with the preceding expression obtained from the specialized Frenkel formula, we find
\begin{align}
    \sum _{L \in P_{++}^{\kappa }} S_{L, \Lambda} N_{L,\kappa }(\tau , v)
    =
\sum_{\lambda \in P_{++}}
\det(\sigma_\lambda) S_{\hat\sigma_\lambda(\lambda), \Lambda}
J_\lambda(v)\,
q^{\frac{\lambda^2}{2\kappa}} .
\end{align}

Multiplying both sides by $\overline{S_{\Lambda,K}}$ and summing over $\Lambda\in P_{++}^{\kappa}$, and using
\begin{align}
\sum_{\Lambda\in P_{++}^{\kappa}}
\overline {S_{L\Lambda}} {S_{\Lambda K}}
= \delta_{LK},
\end{align}
we get
\begin{align}
N_{K,\kappa }(\tau , v)
    =
\sum_{\lambda \in P_{++}}
\det(\sigma_\lambda)\,
\delta_{\hat\sigma_\lambda(\lambda),K}
J_\lambda(v)\,
q^{\frac{\lambda^2}{2\kappa}}, \quad K \in P^\kappa_{++}.
\end{align}
Shifting $\lambda\mapsto\lambda+\rho$, we can recast this expression in a form adapted to the heat equation on $G$.
Writing $\sigma_{\lambda+\rho}=s_\lambda$ and using the shifted Weyl action $s\circ\lambda=s(\lambda+\rho)-\rho$, the corresponding shifted affine action is defined by $\hat s\circ\lambda =s\circ\lambda +\kappa\gamma_{\lambda+\rho}$. In terms of the Weyl-invariant Kac numerator~\eqref{eq:WeylInvKacNum}, the result becomes the Weyl-invariant character expansion
\begin{align}
\mathcal{N}_{K+\rho, k+h^\vee}(\tau, v) 
    = 
\sum_{\lambda \in P_{+}}
\det(s_{\lambda})\,
\delta_{\hat s_{\lambda}\circ \lambda,\,K}\,
\chi_{\lambda}(v)\,
q^{\frac{c_2(\lambda)}{2(k+h^{\vee})}},\, \quad K \in P^k_{+}.
\end{align}
The Casimir dependence makes the heat-equation structure manifest:
\begin{align}
\frac{1}{2\pi i}\frac{\partial}{\partial\tau}\mathcal\, \mathcal N_{K+\rho,k+h^\vee}
 =
\frac{1}{2(k+h^\vee)}\Delta_G\mathcal \, \mathcal N_{K+\rho,k+h^\vee}.
\end{align}
At $k=0$ and $K=0$, this reduces to the normalized affine Weyl denominator, which admits an infinite product expansion,
\begin{align}
   \mathcal{M}(\tau, v) = \mathcal{N}_{\rho,h^\vee} =q^{-\frac{(\rho,\rho)}{2h^{\vee}}} \frac{N_{\rho,\,h^\vee }(\tau , v)}{J_\rho(v)} = (q,q)_\infty^r \prod_{\alpha\in R} (e^{2\pi i (v,\alpha)} q,q)_\infty,
\end{align}
where 
\begin{align}
    (z ,q)_\infty=\prod_{n=0}^{\infty} (1-q^n z),
\end{align}
is the $q$-Pochhammer symbol. This is the Macdonald form of the affine Weyl denominator identity~\cite{macdonald72, Lu:2025ecb}.

\section{Summary and discussion} \label{sec:summary}
In this work, we have studied the localization formula for the torus partition
function of WZW models with compact, connected, and simply connected simple Lie
groups, based on the supersymmetric localization approach to SWZW
models~\cite{murthyW25}. We identified an additional phase factor in the
localization formula originating from the global holonomy of the WZ term, in
agreement with the exact result obtained from the Hamiltonian formulation based
on affine Kac--Moody algebras. Upon abelianization, this phase is naturally
interpreted as a flat Kalb--Ramond $B$--field coupling in the corresponding
Narain description. At the same time, the torus partition functions of both WZW
and SWZW models admit a transparent description in terms of Siegel--Narain theta
functions, providing a direct connection between the global topology of the WZ
term and the Narain lattice data of the abelianized theory. We have also
recovered Frenkel's formula in the quantum-mechanical particle limit and
elucidated its relation to the Weyl--Kac character formulas.

These results suggest that supersymmetric localization may be viewed as part of
a broader localization framework connecting WZW models, quantum mechanics on
group manifolds, and their representation-theoretic descriptions. A natural
first direction is therefore to clarify the relation between the supersymmetric
localization formula derived here and the Duistermaat--Heckman (DH) equivariant
localization approach~\cite{Wendt2001}. The space of maps from $T^2$ to $G$, or
to its complexification $G_{\mathbb C}$, is naturally related to the double loop
group, which admits a one-dimensional central extension
$\widetilde{LLG}_{\mathbb C}$~\cite{EtingofFrenkel1992current}. A relevant
coadjoint orbit can be identified with
$\widetilde{LLG}_{\mathbb C}/T_{\mathbb C}$ and carries a nondegenerate
symplectic form. The action of $S^1\times S^1\times T$, generated by loop
rotations and conjugation, is Hamiltonian, with the gauged WZW action providing
the corresponding Hamiltonian. The WZW torus partition function can thus be
viewed as a DH integral associated with this symplectic data.

In this DH description, the fixed-point locus of the
$S^1\times S^1\times T$ action is labeled by
$W\times Q^\vee\times Q^\vee$, while the Weyl-group sign arises from the
character of the torus action on the tangent space at the localization locus.
This leads to a localization formula organized as a sum over
$W\times Q^\vee\times Q^\vee$, closely paralleling the formula obtained from
supersymmetric localization~\cite{Wendt2001}. Similarly, DH localization
underlies the derivations of both the Weyl character and Weyl--Kac formulas,
through path-integral representations of group quantum mechanics and chiral WZW
models with Wilson line insertions~\cite{perret90}, respectively. Through the
Peter--Weyl theorem, these formulas are in turn closely related to Frenkel's
formula and to the WZW torus partition function. The close parallel among these
constructions suggests that DH equivariant localization, supersymmetric
localization, and the representation-theoretic formulas may be different
realizations of a common underlying localization framework.

It would therefore be interesting to develop this framework systematically and
to establish a precise relation between DH and supersymmetric localization. Such
a relation could provide a geometric interpretation of the localization formula
derived in this work and make manifest the connections among the Weyl and affine
character formulas, the WZW partition function, and their quantum-mechanical
limits. In particular, it would be interesting to understand geometrically how
the global WZ phase identified here enters the symplectic data underlying the DH
description, and whether the agreement between the two localization formulas can
be established directly at the level of their fixed-point contributions.

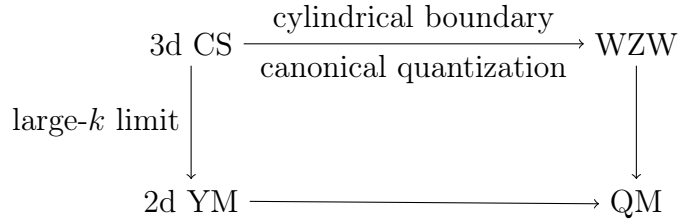
\begin{figure}[H]
\centering
\begin{tikzpicture}[
node distance=1.5cm and 4.5cm,
every node/.style={align=center}
]
\node (CS) {3d CS};
\node (WZW) [right=of CS] {WZW};
\node (YM) [below=of CS] {2d YM};
\node (QM) [below=of WZW] {QM};

\path[->]
(CS) edge node[midway, above] {cylindrical boundary}
          node[midway, below] {canonical quantization} (WZW)
(CS) edge node[midway, left] {large-$k$ limit} (YM)
(WZW) edge node[midway, right] {} (QM)
(YM) edge node[midway, below] {} (QM);

\end{tikzpicture}
\caption{A schematic network relating Chern--Simons theory, WZW theory,
2d Yang--Mills theory, and quantum mechanics on a group manifold.}
\label{fig:duality-web}
\end{figure}

This localization perspective may also be placed in the broader network relating
quantum mechanics on a group manifold, WZW theory, two-dimensional Yang--Mills
theory~\cite{witten91,witten92}, and three-dimensional Chern--Simons
theory~\cite{Witten1988CSJones,elitzurMooreSeibergSchwimmer89}, as illustrated
schematically in Figure~\ref{fig:duality-web}.
Within this broader network, two-dimensional Yang--Mills theory and
three-dimensional Chern--Simons theory can be formulated on $S^1\times I$ and
$T^2\times I$, respectively, and their Hamiltonian descriptions are closely
related to the representation-theoretic structures appearing in group-manifold
quantum mechanics and WZW theory. Through the cylinder interpretation,
partition functions with boundary conditions are related to the corresponding
Hamiltonian propagators. Both two-dimensional Yang--Mills and three-dimensional
Chern--Simons theory also admit non-abelian versions of the DH
approach~\cite{witten92,BeasleyWitten2005}. It would therefore be interesting to
investigate how supersymmetric localization, DH localization, and the
corresponding Hamiltonian descriptions fit together across this network, and in
particular to clarify the relation among the localization descriptions of
group-manifold quantum mechanics, WZW theory, and chiral WZW theory.

From the perspective of the abelianized theory, a complementary direction is to
understand how this localization framework interacts with its moduli and duality
structure. The Narain description obtained here selects a distinguished point in
the moduli space of toroidal CFTs, suggesting that one may study how the
localization formula behaves under exactly marginal current--current
deformations of WZW models~\cite{Forste:2003km} and under $T$-duality
transformations~\cite{Kiritsis:1993ju,Gaberdiel:1995mx}. More broadly, it would
be interesting to explore whether localization provides a useful perspective on
averaging over the resulting moduli
spaces~\cite{Maloney:2020nni,Afkhami-Jeddi:2020ezh,Dong:2021wot} and on special
Narain theories such as code CFTs~\cite{Dymarsky:2020qom}.

Beyond these broader directions, a more immediate extension of the present
analysis is to WZW models with non-simply connected target groups. In this case,
additional global topological data enter both the WZ amplitude and the
localization formula. We study this extension separately
in~\cite{luWZWFGK2026}. We find that these phases, together with the
corresponding fermion global anomaly phases in the supersymmetric theory,
reproduce the simple-current orbifold result. Upon abelianization, the resulting
localization formula again admits a shifted Narain lattice description refined
by orbifold discrete torsion, extending the relation among the global geometry
of the WZ term, abelianization, and affine representation theory established
here.

Finally, beyond this immediate extension, it would be interesting to explore
more general topological aspects of both the target space and the worldsheet.
On the target-space side, this includes WZW models with disconnected or
non-compact groups and, more generally, the role of global WZ and gerbe data in
localization. On the worldsheet side, a natural direction is to consider
higher-genus Riemann surfaces, where the larger set of nontrivial cycles should
lead to a richer interplay between WZ holonomies, winding sectors, and
localization data. Another important extension is to worldsheets with
boundaries, and hence to open strings and D-branes on group
manifolds~\cite{AlekseevSchomerus1999,GawedzkiReis2002}. In this setting, the
global WZ and gerbe data are closely related to the geometry and charges of WZW
D-branes~\cite{GawedzkiReis2002}, providing a natural point of contact with
twisted $K$-theory~\cite{FreedHopkinsTelemanI,FreedHopkinsTelemanII,
FreedHopkinsTelemanIII}. It would be interesting to understand how these
structures are realized in the localization framework and how they appear after
abelianization. These generalizations may provide a broader setting in which to
understand how the topology of the target space and the worldsheet, as well as
boundary data, is encoded in the localized theory. We leave these questions for
future work.

\appendix

\section{Gram matrix of the simple coroots}
\label{apx:gramMatrix}

For completeness, we summarize the parity properties of the Gram matrix of the simple coroots for the irreducible Dynkin types. We use the normalization $(\theta,\theta)=2$, where $\theta$ is the highest root. The purpose is to verify that, apart from the $C_n$ case discussed in the main text, every irreducible type has an odd entry in its coroot Gram matrix.

For the simply-laced algebras
$A_n$, $D_n$, $E_6$, $E_7$, and $E_8$, the roots and coroots coincide.
Consequently,
\begin{align}
    \mathsf C(G)_{ij}
    =
    2\delta_{ij}-\delta_{i\sim j},
    \qquad
    G=A_n,D_n,E_6,E_7,E_8,
\end{align}
where $i\sim j$ means that the corresponding nodes of the Dynkin diagram
$\Gamma_G$ are connected by an edge. Equivalently,
\begin{align}
    \mathsf C(G)=2I-{\rm Adj}(\Gamma_G),
\end{align}
where ${\rm Adj}(\Gamma_G)$ denotes the adjacency matrix of $\Gamma_G$.
Thus every edge of the Dynkin diagram gives an off-diagonal entry $-1$,
and hence $\mathsf C(G)$ is not entrywise even.

For $B_n$, with the standard numbering,
\begin{align}
    \mathsf C(B_n)_{ij}
=
2\delta_{ij}
-\delta_{i,j+1}
-\delta_{i,j-1}
+2\delta_{i,n}\delta_{j,n},
\qquad n\geq2.
\end{align}
Thus
\begin{align}
    \mathsf C(B_n)
=
\begin{pmatrix}
2&-1&0&\cdots&0\\
-1&2&-1&\ddots&\vdots\\
0&-1&2&\ddots&0\\
\vdots&\ddots&\ddots&2&-1\\
0&\cdots&0&-1&4
\end{pmatrix},
\end{align}
which again contains the odd entries $-1$.

For $F_4$ and $G_2$, one finds
\begin{align}
    \mathsf C(F_4)
    &=
    \begin{pmatrix}
    2&-1&0&0\\
    -1&2&-2&0\\
    0&-2&4&-2\\
    0&0&-2&4
    \end{pmatrix},
    \\
    \mathsf C(G_2)
    &=
    \begin{pmatrix}
    6&-3\\
    -3&2
    \end{pmatrix}.
\end{align}
Thus both matrices contain odd off-diagonal entries.

It follows that, among all irreducible Dynkin types,
\begin{align}
    \mathsf C(G)\ \text{is entrywise even}
    \quad\Longleftrightarrow\quad
    G\ \text{is of type }C_n.
\end{align}
Equivalently, the coroot lattice $Q^\vee$ has an even bilinear pairing,
\begin{align}
    (m,w)\in2\mathbb Z
    \qquad
    \text{for all }m,w\in Q^\vee,
\end{align}
precisely for type $C_n$.

\section{Localization equation and holomorphic adjoint bundles}
\label{apx:bundleWZW}

Let $\Sigma$ be a Riemann surface, and let $P_L$ and $ P_R $ be principal bundles with structure groups $G_L$ and $G_R$, equipped with connections $A^L$, $ A^R$ respectively. We specialize below to the case in which $P_L$ and $P_R$ are trivial and the background connections $A^L$, $A^R$ are flat.
The WZW field is a section of the associated bundle
\begin{align}
\mathcal E 
=
(P_L \times P_R)
\times_{(G_L \times G_R)}
G,
\end{align}
where $G_L \times G_R$ acts on the fiber $G$ by
\begin{align}
(h_L,h_R)\cdot g
=
h_L g h_R^{-1}.
\end{align}

In a local trivialization of $P_L$ and $P_R$, the section $g$ is represented by an ordinary $G$-valued function $g(z,\bar z)$, and a change of trivialization gives the familiar gauge transformation law
\begin{align}
g
\mapsto
h_L g h_R^{-1}.
\end{align}

Consider the adjoint bundles ${\rm Ad}(P_L) = P_L \times_{G_L} \mathfrak g$ and ${\rm Ad}(P_R)= P_R \times_{G_R} \mathfrak g$. A section $g$ of $\mathcal{E}$ then induces, fiberwise, a map
\begin{align}
{\rm Ad}_g:
{\rm Ad}(P_R)
&\to
{\rm Ad}(P_L),
\end{align}
which in a local trivialization is given by
\begin{align}
X_R
\mapsto
gX_Rg^{-1}.
\end{align}

To see that this is globally well defined, consider a change of local trivialization,
\begin{align}
g
\mapsto
h_L g h_R^{-1}, \qquad 
X_R
\mapsto
h_R X_R h_R^{-1}.
\end{align}
Then
\begin{align}
gX_Rg^{-1}
&\mapsto
(h_Lgh_R^{-1})
(h_RX_Rh_R^{-1})
(h_Rg^{-1}h_L^{-1})
\\
&=
h_L(gX_Rg^{-1})h_L^{-1}.
\end{align}
Thus the transformed expression has precisely the transition law for a section of ${\rm Ad}(P_L)$.

The $(0,1)$ components of the background connections define $(0,1)$-connections on the associated complex vector bundles. On the adjoint bundles ${\rm Ad}(P_R)
$ and $
{\rm Ad}(P_L)$, these take the form
\begin{align}
\bar\partial_R
&=
\bar\partial
+
{\rm ad }( A^{R}_{(0,1)}),
&
\bar\partial_L
&=
\bar\partial
+
{\rm ad }( A^{L}_{(0,1)}).
\end{align}
For an underlying representation of the gauge group, the corresponding operators are
\begin{align}
D^R_{\bar z}
&=
\partial_{\bar z}+A^R_{\bar z},
&
D^L_{\bar z}
&=
\partial_{\bar z}+ A^L_{\bar z},
\end{align}
with the gauge fields understood in the appropriate representation.

A $(0,1)$-connection defines a holomorphic structure precisely when it is integrable, namely when $\bar\partial^2=F^{0,2}=0$. Since $\Sigma$ is a Riemann surface, $\Omega^{0,2}(\Sigma)=0$, so this condition is automatic. Thus the $(0,1)$ components of $A^L$ and $A^R$ define holomorphic structures on the corresponding complexified adjoint bundles.
Hence the localization equations can naturally be interpreted in the category of holomorphic vector bundles.

The localization equation determines $g$ through
\begin{align}
A^R_{\bar z} = 
g^{-1} A^L_{\bar z}g
+ g^{-1} \partial_{\bar z}g ,
\end{align}
This further implies the intertwining relation 
\begin{align}
D^L_{\bar z}( g s)
=
g(D^R_{\bar z}s).
\end{align}
for any local section $s$.
Thus the central geometric interpretation of the localization equation is that $g$ intertwines the two $(0,1)$-connections. In particular, the section $g$ induces a fiberwise invertible map
\begin{align}
{\rm Ad}_g:
({\rm Ad}_{\mathbb{C}}(P_R),\bar\partial_R)
\to 
({\rm Ad}_{\mathbb{C}}(P_L),\bar\partial_L),
\end{align}
satisfying
\begin{align}
     \bar\partial_L\circ {\rm Ad}_g
 =
{\rm Ad}_g\circ\bar\partial_R.
\end{align}

For a general section $g\in\Gamma(\mathcal E)$, ${\rm Ad}_g$ is only a smooth bundle isomorphism; the localization equation is precisely the additional condition that this isomorphism be holomorphic.
Hence the localization equation is not merely a local differential equation for a matrix-valued field. It is the condition that the WZW field identify the two holomorphic structures.

The relation 
\begin{align}
A^R_{\bar z} = 
g^{-1} A^L_{\bar z}g
+ g^{-1} \partial_{\bar z}g = (A^L_{\bar z})_g,
\end{align}
takes exactly the form of usual gauge-transformation law for a connection.
It is therefore natural to refer to $g$ locally as a gauge transformation. Globally, however, $P_L$ and $P_R$ need not be the same principal bundle. Consequently, $g$ should not automatically be regarded as a globally defined gauge transformation of a single principal bundle. Instead, it should be regarded as a smooth isomorphism between the adjoint bundles, which becomes a holomorphic isomorphism when the localization equation is satisfied.
Thus one obtains the geometric equivalence
\begin{align}
\begin{array}{c}
g\in\Gamma(\mathcal E)\text{ solves }\\ \text {the localization equation}
\end{array}
\quad\Longleftrightarrow\quad
\begin{array}{c}
{\rm Ad}_g\text{ is a holomorphic}\\
\text{isomorphism of the two adjoint bundles}.
\end{array}
\end{align}
This perspective allows the problem of solving the localization differential equation for $g$ to be recast, when appropriate, as a problem concerning holomorphic vector bundles and their isomorphisms.

This bundle-theoretic description becomes particularly useful after abelianization. We choose a maximal torus $T\subset G$ and denote its normalizer by $N_G(T)$, the Weyl group is given by $W=N_G(T)/T$.  We now specialize further to the abelianized localization locus, where $A^L$ and $A^R$ are flat and Cartan valued.
\begin{align}
A^L_{\bar z},
\,
A^R_{\bar z}
\in
\mathfrak t_{\mathbb C}.
\end{align}
The complexified Lie algebra then decomposes according to the root decomposition,
\begin{align}
\mathfrak g_{\mathbb C}
=
\mathfrak t_{\mathbb C}
\oplus
\bigoplus_{\alpha\in\Delta}
\mathfrak g_\alpha.
\end{align}

Accordingly, the complexified adjoint bundle decomposes into a Cartan sector and root sectors,
\begin{align}
{\rm Ad}_{\mathbb{C}}(P_{L,R})
=
\mathcal{O}^{\oplus r}
\oplus
\bigoplus_{\alpha\in\Delta}
\mathcal{L}_\alpha^{L,R},
\end{align}
where each $\mathcal{L}_\alpha$ is the holomorphic line bundle associated with the root $\alpha$.

The intertwining condition induced by $g$ can consequently be analyzed in terms of the individual root sectors. The localization solutions preserve the Cartan decomposition up to a Weyl transformation $\sigma\in W$. The root sectors are therefore paired according to
\begin{align}
 \mathcal L^R_\alpha
 \simeq
 \mathcal L^L_{\sigma(\alpha)}.
\end{align}
Thus the localization problem decomposes into holomorphic isomorphism problems for the corresponding root line bundles. The localization solution takes the form
\begin{align}
 g=g_\sigma\tilde g,
 \qquad
 \sigma\in W,
 \qquad
 \tilde g:\Sigma\to T,
\end{align}
where $g_\sigma\in N_G(T)$ is a representative of the Weyl element $\sigma\in W$. For instance,
\begin{align}
    g_{\sigma} = \{ \begin{pmatrix}
        1 & 0\\
        0 & 1
    \end{pmatrix}, \begin{pmatrix}
        0 & -1\\
        1 & 0
    \end{pmatrix}\}
\end{align}
for $G = SU(2)$.

\section{WZ amplitude and $B$-field amplitude}
\label{apx:WZ&Bfield}

We first recall some basic properties of the WZ functional and its
relation to the Polyakov--Wiegmann (PW) cocycle. The WZ functional is
\begin{align}
\Gamma_{\rm WZ}(g)
=
\frac{1}{12\pi}
\int_B
\tr\left[
(g^{-1}dg)^3
\right],
\end{align}
and the corresponding WZ amplitude on $T^2$ at level $k\in\mathbb{Z}$ is
\begin{align}
A_{\rm WZ}(g)
=
e^{-ik\Gamma_{\rm WZ}(g)}.
\end{align}

The WZ functional is not additive under pointwise multiplication of
group-valued fields. For two maps $g,h:T^2\to G$, the Polyakov--Wiegmann
(PW) formula takes the form~\cite{PolyakovWiegmann1984}
\begin{align}
\label{eq:PWformula}
\Gamma_{\rm WZ}(gh)
=
\Gamma_{\rm WZ}(g)+\Gamma_{\rm WZ}(h)+\phi_{\rm WZ}(g,h),
\end{align}
where
\begin{align}
\phi_{\rm WZ}(g,h)
=
-\frac{1}{4\pi}
\int_{T^2}
\tr\left[
g^{-1}dg\wedge(dh)h^{-1}
\right].
\end{align}
Thus, the failure of $\Gamma_{\rm WZ}$ to be additive is captured by the local
two-dimensional functional $\phi(g,h)_{\rm WZ}$. Upon exponentiation, this gives a
$U(1)$-valued factor
\begin{align}
\Phi_{\rm WZ}(g,h)
=
e^{-ik\phi_{\rm WZ}(g,h)},
\end{align}
such that
\begin{align}
A_{\rm WZ}(gh)
=
A_{\rm WZ}(g)A_{\rm WZ}(h)\Phi_{\rm WZ}(g,h).
\end{align}
Associativity of pointwise group multiplication implies the corresponding
2-cocycle condition for $\Phi_{\rm WZ}$. We refer to $\Phi_{\rm WZ}$ as the PW cocycle at
level $k$.

These data admit a geometric interpretation in terms of the basic bundle
gerbe with connection $\mathcal{G}_k$ on $G$. Its curvature is the closed
three-form
\begin{align}
H_k
=
\frac{k}{24\pi^2} \tr(g^{-1}dg)^3,
\end{align}
Since $H_k$ is in general nontrivial in
cohomology, the WZ term cannot be described globally by a single two-form
$B$. Instead, it is described by local $B$-fields together with their
transition data, encoded by the bundle gerbe $\mathcal{G}_k$~\cite{Waldorf2008thesisGerbe}. Its
$U(1)$-valued holonomy on the worldsheet is precisely the WZ amplitude,
\begin{align}
\mathcal{H}_{\mathcal{G}_k}(g)
=
A_{\rm WZ}(g).
\end{align}

It is useful to compare this nonabelian construction with its abelian
counterpart. In particular, upon restricting the WZ data to an abelian
subgroup, such as a maximal torus $T\subset G$, the invariant three-form
vanishes. The resulting gerbe data can therefore be described in terms of
a globally defined closed two-form. We thus consider a constant KR $B$-field
satisfying
\begin{align}
H=dB=0.
\end{align}
The corresponding worldsheet coupling is
\begin{align}
\Gamma_{\rm KR}(X) = \frac{1}{8\pi}
\int_{T^2}
B_{ij}\,dX^i\wedge dX^j,
\end{align}
and its holonomy is
\begin{align}
A_{\rm KR}(X)
=
e^{
-i \Gamma_{\rm KR}(X)
}.
\end{align}

The $B$-field satisfies an abelian
analogue of the PW formula. For two maps
$X,Y:T^2\to T$, one finds
\begin{align}
A_{\rm KR}(X+Y)
=
A_{\rm KR}(X)A_{\rm KR}(Y) \Phi_{\rm KR}(X,Y)
,
\end{align}
where
\begin{align}
    \Phi_{\rm KR}(X,Y)=
\exp\{
-\frac{i}{4\pi}
\int_{T^2}
B_{ij}\,dX^i\wedge dY^j
\}.
\end{align}
Thus, just as the nonabelian WZ amplitude is multiplicative up to the PW
cocycle, the abelian $B$-field holonomy is additive up to a $U(1)$-valued
two-cocycle.

Consider a toroidal sigma model with target $T^d=\mathbb{R}^d/(2\pi L)$ where $L$ is the period lattice. A map $X:T^2\to T^d$ is characterized,
up to its single-valued part, by its winding data,
\begin{align}
dX
=
m\,dt+w\,ds,
\qquad
m,w\in L.
\end{align}
For a constant $B$-field, the holonomy depends only on the corresponding
cohomology class of $dX$ and is given by
\begin{align}
A_{\rm KR}(X)
=
\exp\left\{
-\pi i\,B_{ij}m^i w^j
\right\}.
\end{align}
Thus, a flat $B$-field assigns a $U(1)$ phase to each winding sector labeled with $m,w \in L$.

This phase can be obtained directly from the abelian
PW formula, applying to $X_{m,w} = X_m + X_w$ with $X_m= mt$ and $
X_w= w s$. Since each of these configurations has vanishing $B$-field holonomy
separately, the PW formula gives
\begin{align}
A_{\rm KR}(X_{m,w})
=
\exp\{
-\frac{i}{4\pi}
\int_{T^2}
B_{ij}\,dX_m^i\wedge dX_w^j
\}
=
\exp\left\{
-\pi i\,B_{ij}m^i w^j
\right\}.
\end{align}
In this sense, the $B$-field amplitude of a classical winding
configuration is precisely the abelian PW cocycle evaluated on the two
elementary winding configurations.

The same interpretation applies to the PW cocycle of the WZ functional
for abelianized configurations. Upon restriction of the WZ data to the
maximal torus, the nonabelian PW formula reduces to the abelian formula
above, and the corresponding cocycle becomes the lattice two-cocycle
determined by the flat $B$-field. Thus, the phase associated with a
winding sector can be viewed as the abelianized PW cocycle of the WZ
amplitude.

Geometrically, this may be viewed as the reduction of the basic gerbe on
$G$ to flat gerbe data on the maximal torus. Equivalently, the holonomy
of the WZ term restricted to $T$ is encoded by Weyl-equivariant
gerbe data, which can in turn be described in terms of a Weyl-equivariant
collection of line bundles over the maximal torus. The resulting lattice
two-cocycle is therefore the abelianized form of the PW cocycle and
provides the natural bridge between the WZ amplitude and the $B$-field
phase factors appearing in the torus winding sectors.

\section{Gauging of WZ term and $B$-field coupling}
\label{apx:WZKRgauging}

The gauging of WZ and $B$-field couplings under a Lie group symmetry of the sigma model admits two complementary frameworks: one based on transgression formulas relating anomaly polynomials to Chern--Simons forms~\cite{Alvarez-GaumeGinsparg1984Anomaly, HullSpence1990GaugedWZ}, and the other based on equivariant cohomology and the Mathai--Quillen formalism~\cite{Witten1991holomorpFac}. The transgression formulation makes the Green--Schwarz interpretation particularly transparent.

The transgression form is~\cite{ManesStoraZumino1985Chiral,
Alvarez-GaumeGinsparg1984Anomaly,MoraOTZ2006Transgression,
Nakahara2003book,MooreSaxena2025TASI}
\begin{align}
T[A_1,A_0]
=
2\int_0^1dt\,
\tr\!\left[(A_1-A_0)F_t\right],
\end{align}
where $A_t=tA_1+(1-t)A_0$ and $F_t$ is its curvature.

For the nonabelian WZ model with $G_L\times G_R$ symmetry, taking
\begin{align}
A_t=A_g^L-t\mathcal J,
\end{align}
gives
\begin{align}
T[A^R,A_g^L]
&=
-2\int_0^1dt\,
\tr\!\left[
\mathcal J
\left(
(1-t)g^{-1}F^Lg+tF^R+(t^2-t)\mathcal J^2
\right)
\right]
\nonumber\\
&=
-\tr\!\left[
\mathcal J(g^{-1}F^Lg+F^R)
\right]
+\frac13\tr\mathcal J^3
\nonumber\\
&=
g^\ast\Omega^A-CS[A^L]+CS[A^R],
\end{align}
where
\begin{align}
CS[A]
=
\tr\left(A\wedge dA+\frac23A^3\right),
\end{align}
and $g^\ast\Omega^A$ is the $G_L\times G_R$-equivariant extension of the
Cartan three-form,
\begin{align}
g^\ast\Omega^A
=
g^\ast\Omega
-
d\left(
A^L\wedge g\,dg^{-1}
-g^{-1}dg\wedge A^R
-A^L\wedge gA^Rg^{-1}
\right).
\end{align}
Consequently,
\begin{align}
\frac{1}{4\pi}dT[A^R,A_g^L]
=
-\frac{1}{4\pi}\tr F_L^2
+\frac{1}{4\pi}\tr F_R^2,
\end{align}
reproducing the perturbative $G_L\times G_R$ anomaly polynomial. The transgression form therefore plays the role of an equivariant Green--Schwarz field strength, with its equivariant completion encoding
the coupling to the background $G_L\times G_R$ gauge fields.

The same construction applies to the abelian theory with
$U(1)^d_L\times U(1)^d_R$ symmetry. Taking
\begin{align}
A_t=A_L+dX-tDX,
\end{align}
gives
\begin{align}
T[A_R,A_L+dX]
&=
-2\int_0^1dt\,
G_{ij}DX^i\wedge
\left((1-t)F_L^j+tF_R^j\right)
\nonumber\\
&=
-G_{ij}DX^i\wedge(F_L^j+F_R^j)
\nonumber\\
&=
d\omega_{\rm eq}-CS[A_L]+CS[A_R],
\end{align}
where
\begin{align}
CS[A]
=
G_{ij}A^i\wedge dA^j,
\qquad
\omega_{\rm eq}
=
-G_{ij}\left[
(A_L^i+A_R^i)\wedge dX^j
-A_L^i\wedge A_R^j
\right].
\end{align}
Here $\omega_{\rm eq}$ provides the local gauge-field-dependent completion
required by the $U(1)^d_L\times U(1)^d_R$ symmetry.
It follows that
\begin{align}
\frac{1}{4\pi }dT[A_R,A_L+dX]
=
-\frac{1}{4\pi }G_{ij}F_L^i\wedge F_L^j
+
\frac{1}{4\pi }G_{ij}F_R^i\wedge F_R^j,
\end{align}
reproducing the perturbative $U(1)^d_L\times U(1)^d_R$ anomaly
polynomial. Thus the abelian transgression likewise has the interpretation of an
equivariant Green--Schwarz field strength, with $\omega_{\rm eq}$ providing
its local equivariant completion.

There is, however, an additional freedom in the abelian case.
The pullback of the Cartan three-form to the maximal torus vanishes, and
the remnant of the WZ coupling may be represented by a globally defined
flat $B$-field. The equivariant $B$-field therefore takes the form
\begin{align}
X^\ast B_{\rm eq}
=
X^\ast B+\omega_{\rm eq},
\qquad
dB=0.
\end{align}
The shift by the flat part $B$ does not modify the transgression form and is therefore invisible to the local anomaly polynomial. It survives instead as global holonomy. Hence in both the nonabelian and abelian descriptions the equivariant completion encodes the coupling to the background gauge fields, while
the abelianized theory admits the additional freedom of a flat $B$-field
shift. This flat ambiguity is the Narain $B$-field modulus and, in the
present abelianization, encodes the global WZ holonomy of the original
nonabelian theory.

\section{Solid-torus extension and the WZ amplitude}
\label{apx:solidTorus}

Here we give an explicit solid-torus extension of the torus-valued
classical configuration
\begin{align}
g_{\rm cl}(s,t)
=
e^{i(mt+ws)},
\qquad
m,w\in Q^\vee,
\end{align}
adapted to the decomposition of the winding data into simple-coroot
directions. This choice also provides a direct evaluation of the WZ
functional through the PW formula.

We first recall an explicit disk filling in $SU(2)$. We use the standard
Pauli-matrix conventions
\begin{align}
\sigma_2
=
\begin{pmatrix}
0&-i\\
i&0
\end{pmatrix},
\qquad
\sigma_3
=
\begin{pmatrix}
1&0\\
0&-1
\end{pmatrix}.
\end{align}
For $n\in\mathbb Z$, define
\begin{align}
f_n:D_s^2\longrightarrow SU(2)
\end{align}
by
\begin{align}
f_n(s,r)
&=
e^{\frac{ins}{2}\sigma_3}
e^{i\frac{\pi}{2}(1-r)\sigma_2}
e^{\frac{ins}{2}\sigma_3}
\nonumber\\
&=
\begin{pmatrix}
\sin\frac{\pi r}{2}\,e^{ins}
&
\cos\frac{\pi r}{2}
\\
-\cos\frac{\pi r}{2}
&
\sin\frac{\pi r}{2}\,e^{-ins}
\end{pmatrix}.
\end{align}
At the boundary $r=1$,
\begin{align}
f_n(s,1)
=
\begin{pmatrix}
e^{ins}&0\\
0&e^{-ins}
\end{pmatrix}
=
e^{ins\sigma_3},
\end{align}
whereas at the center $r=0$ the map is independent of $s$. Thus $f_n$
provides an explicit disk filling of the loop $e^{ins\sigma_3}$.

We now apply this construction to the simple-coroot directions of $G$.
Write
\begin{align}
m=\sum_{i=1}^r m_i\alpha_i^\vee,
\qquad
w=\sum_{i=1}^r w_i\alpha_i^\vee,
\qquad
m_i,w_i\in\mathbb Z.
\end{align}
For each simple coroot $\alpha_i^\vee$, let
\begin{align}
\iota_i:SU(2)\hookrightarrow G
\end{align}
denote the corresponding root embedding, normalized by
\begin{align}
\iota_i\left(e^{i\theta\sigma_3}\right)
=
e^{i\theta\alpha_i^\vee}.
\end{align}
The corresponding embedded disk filling is
\begin{align}
f^{(i)}_{w_i}(s,r)
&=
\iota_i\left(f_{w_i}(s,r)\right)
\nonumber\\
&=
\iota_i\left(
e^{\frac{iw_i s}{2}\sigma_3}
e^{i\frac{\pi}{2}(1-r)\sigma_2}
e^{\frac{iw_i s}{2}\sigma_3}
\right),
\end{align}
and satisfies
\begin{align}
f^{(i)}_{w_i}(s,1)
=
e^{iw_i s\alpha_i^\vee}.
\end{align}

To incorporate the winding along the $t$-cycle, define
\begin{align}
h_i(s,t,r)
=
f^{(i)}_{w_i}(s,r)
e^{itm_i\alpha_i^\vee}.
\end{align}
We then choose the solid-torus extension
\begin{align}
\widetilde g_{\rm cl}(s,t,r)
=
h_1(s,t,r)h_2(s,t,r)\cdots h_r(s,t,r).
\end{align}
On the boundary $r=1$, all factors lie in the maximal torus and hence
commute. Therefore
\begin{align}
\widetilde g_{\rm cl}(s,t,1)
=
\prod_i
e^{i(sw_i+tm_i)\alpha_i^\vee}
=
e^{i(ws+mt)}
=
g_{\rm cl}(s,t).
\end{align}
Thus $\widetilde g_{\rm cl}$ provides an explicit extension of
$g_{\rm cl}$ to $D_s^2\times S_t^1$.

We can now evaluate its WZ functional by repeated application of the
PW formula. On the boundary the factors $h_i$ are Cartan-valued and
commute. Their Maurer--Cartan forms are therefore additive under
multiplication, so the successive PW cross terms reduce to a sum of
pairwise contributions:
\begin{align}
\Gamma_{\rm WZ}(\widetilde g_{\rm cl})
=
\sum_i\Gamma_{\rm WZ}(h_i)
+
\sum_{i<j}\phi_{\rm WZ}(h_i,h_j)
\pmod{2\pi}.
\end{align}

Recall the Gram matrix of the simple coroots,
$\mathsf C_{ij}=(\alpha_i^\vee,\alpha_j^\vee)$,
defined in~\eqref{eq:GramMatrix}. The diagonal contributions are
determined entirely within the individual embedded $SU(2)_i$ subgroups:
\begin{align}
\Gamma_{\rm WZ}(h_i)
=
\pi\,\mathsf C_{ii}m_iw_i
\pmod{2\pi}.
\end{align}
For $i\neq j$, the PW cocycle is determined by the boundary values.
Since
\begin{align}
h_i^{-1}dh_i\big|_{\partial}
=
i\alpha_i^\vee
(m_i\,dt+w_i\,ds),
\end{align}
the PW formula gives
\begin{align}
\phi_{\rm WZ}(h_i,h_j)
=
\pi\mathsf C_{ij}
(m_iw_j-m_jw_i)
\pmod{2\pi}.
\end{align}
Combining the diagonal and off-diagonal contributions yields
\begin{align}
\Gamma_{\rm WZ}(\widetilde g_{\rm cl})
&=
\pi\sum_i\mathsf C_{ii}m_iw_i
+
\pi\sum_{i<j}\mathsf C_{ij}
(m_iw_j-m_jw_i)
\pmod{2\pi}.
\end{align}

To compare this with the invariant Cartan pairing, note that
\begin{align}
\pi(m,w)
=
\pi\sum_i\mathsf C_{ii}m_iw_i
+
\pi\sum_{i<j}\mathsf C_{ij}
(m_iw_j+m_jw_i).
\end{align}
The difference is
\begin{align}
2\pi\sum_{i<j}
\mathsf C_{ij}m_jw_i
\in2\pi\mathbb Z,
\end{align}
since $\mathsf C_{ij},m_i,w_i\in\mathbb Z$. Consequently,
\begin{align}
\Gamma_{\rm WZ}(g_{\rm cl})
=
\pi(m,w)
\pmod{2\pi}.
\end{align}

Thus the ordered PW decomposition naturally produces an antisymmetric
representative of the coroot pairing, which is equivalent modulo $2$
to the symmetric invariant pairing $(m,w)$. This is precisely the same
mod-$2$ structure that appears in the flat $B$-field
representative~\eqref{eq:Brepresentative} of the abelianized theory.

\bibliographystyle{myJHEP}
\bibliography{mybib}

\providecommand{\href}[2]{#2}\begingroup\raggedright\begin{thebibliography}{10}

\bibitem{Witten1983WZW}
E.~Witten, { \it {Nonabelian Bosonization in Two-Dimensions}},   {\rm Commun. Math. Phys.} {\bf 92} (1984) 455--472.

\bibitem{KnizhnikZamolodchikov1984WZW}
V.~G. Knizhnik and A.~B. Zamolodchikov, { \it {Current Algebra and Wess-Zumino Model in Two-Dimensions}},   {\rm Nucl. Phys. B} {\bf 247} (1984) 83--103.

\bibitem{gepnerWitten86}
D.~Gepner and E.~Witten, { \it {String Theory on Group Manifolds}},   {\rm Nucl. Phys. B} {\bf 278} (1986) 493--549.

\bibitem{kac3rd}
V.~G. Kac,  {\rm Infinite Dimensional Lie Algebras}.
\newblock Cambridge University Press, Cambridge, 3~ed., 1990.

\bibitem{Wesszumino1971}
J.~Wess and B.~Zumino, { \it {Consequences of anomalous Ward identities}},   {\rm Phys. Lett. B} {\bf 37} (1971) 95--97.

\bibitem{murthyW25}
S.~Murthy and E.~Witten, { \it {Localization of Strings on Group Manifolds}},   {\rm Commun. Math. Phys.} {\bf 407} (2026), no.~7 154, [\href{http://arxiv.org/abs/2506.20028}{{\tt arXiv:2506.20028}}].

\bibitem{DiVecchiaKnizhnikPR1984SWZW}
P.~Di~Vecchia, V.~G. Knizhnik, J.~L. Petersen, and P.~Rossi, { \it {A Supersymmetric Wess-Zumino Lagrangian in Two-Dimensions}},   {\rm Nucl. Phys. B} {\bf 253} (1985) 701--726.

\bibitem{GawedzkiReis2002braneGerbe}
K.~Gawedzki and N.~Reis, { \it {WZW branes and gerbes}},   {\rm Rev. Math. Phys.} {\bf 14} (2002) 1281--1334, [\href{http://arxiv.org/abs/hep-th/0205233}{{\tt hep-th/0205233}}].

\bibitem{Narain:1986am}
K.~S. Narain, M.~H. Sarmadi, and E.~Witten, { \it {A Note on Toroidal Compactification of Heterotic String Theory}},   {\rm Nucl. Phys. B} {\bf 279} (1987) 369--379.

\bibitem{choiT25}
C.~Choi and L.~A. Takhtajan, { \it {Supersymmetry and trace formulas. Part III. Frenkel trace formula}},   {\rm JHEP} {\bf 06} (2026) 200, [\href{http://arxiv.org/abs/2502.10210}{{\tt arXiv:2502.10210}}].

\bibitem{Frenkel1984}
I.~B. Frenkel, { \it Orbital theory for affine {L}ie algebras},   {\rm Inventiones Mathematicae} {\bf 77} (1984), no.~2 301--352.

\bibitem{Bernard:1987df}
D.~Bernard, { \it {On the Wess-Zumino-Witten Models on the Torus}},   {\rm Nucl. Phys. B} {\bf 303} (1988) 77--93.

\bibitem{Zhao:2025hen}
B.~Zhao, { \it {WZW Partition Functions from Supersymmetric Localization}},  \href{http://arxiv.org/abs/2507.11673}{{\tt arXiv:2507.11673}}.

\bibitem{Witten1991holomorpFac}
E.~Witten, { \it {On Holomorphic factorization of WZW and coset models}},   {\rm Commun. Math. Phys.} {\bf 144} (1992) 189--212.

\bibitem{blauT93}
M.~Blau and G.~Thompson, { \it {Derivation of the Verlinde formula from Chern-Simons theory and the G/G model}},   {\rm Nucl. Phys. B} {\bf 408} (1993) 345--390, [\href{http://arxiv.org/abs/hep-th/9305010}{{\tt hep-th/9305010}}].

\bibitem{Blau:1994rk}
M.~Blau and G.~Thompson, { \it {On diagonalization in map(M,G)}},   {\rm Commun. Math. Phys.} {\bf 171} (1995) 639--660, [\href{http://arxiv.org/abs/hep-th/9402097}{{\tt hep-th/9402097}}].

\bibitem{blauT93lecture}
M.~Blau and G.~Thompson, { \it {Lectures on 2-d gauge theories: Topological aspects and path integral techniques}},  in  {\rm {Summer School in High-energy Physics and Cosmology (Includes Workshop on Strings, Gravity, and Related Topics 29-30 Jul 1993)}}, pp.~0175--244, 10, 1993.
\newblock \href{http://arxiv.org/abs/hep-th/9310144}{{\tt hep-th/9310144}}.

\bibitem{macdonald72}
I.~G. Macdonald, { \it Affine root systems and {Dedekind}'s $\eta$-function},   {\rm Inventiones mathematicae} {\bf 15} (June, 1972) 91--143.

\bibitem{Lu:2025ecb}
Y.~L{\"u}, { \it {Macdonald Identities and Exact Formulas for Superconformal Indices in Super Yang-Mills Theories}},  \href{http://arxiv.org/abs/2511.08468}{{\tt arXiv:2511.08468}}.

\bibitem{Wendt2001}
R.~Wendt, { \it A symplectic approach to certain functional integrals and partition functions},   {\rm Journal of Geometry and Physics} {\bf 40} (2001), no.~1 65--99.

\bibitem{EtingofFrenkel1992current}
P.~I. Etingof and I.~B. Frenkel, { \it {Central extensions of current groups in two-dimensions}},   {\rm Commun. Math. Phys.} {\bf 165} (1994) 429--444, [\href{http://arxiv.org/abs/hep-th/9303047}{{\tt hep-th/9303047}}].

\bibitem{perret90}
R.~E. Perret, { \it {Path integral derivation of characters for Kac-Moody groups}},   {\rm Nucl. Phys. B} {\bf 356} (1991) 229--244.

\bibitem{witten91}
E.~Witten, { \it {On quantum gauge theories in two-dimensions}},   {\rm Commun. Math. Phys.} {\bf 141} (1991) 153--209.

\bibitem{witten92}
E.~Witten, { \it {Two-dimensional gauge theories revisited}},   {\rm J. Geom. Phys.} {\bf 9} (1992) 303--368, [\href{http://arxiv.org/abs/hep-th/9204083}{{\tt hep-th/9204083}}].

\bibitem{Witten1988CSJones}
E.~Witten, { \it {Quantum Field Theory and the Jones Polynomial}},   {\rm Commun. Math. Phys.} {\bf 121} (1989) 351--399.

\bibitem{elitzurMooreSeibergSchwimmer89}
S.~Elitzur, G.~W. Moore, A.~Schwimmer, and N.~Seiberg, { \it {Remarks on the Canonical Quantization of the Chern-Simons-Witten Theory}},   {\rm Nucl. Phys. B} {\bf 326} (1989) 108--134.

\bibitem{BeasleyWitten2005}
C.~Beasley and E.~Witten, { \it {Non-Abelian localization for Chern-Simons theory}},   {\rm J. Diff. Geom.} {\bf 70} (2005), no.~2 183--323, [\href{http://arxiv.org/abs/hep-th/0503126}{{\tt hep-th/0503126}}].

\bibitem{Forste:2003km}
S.~F{\"o}rste and D.~Roggenkamp, { \it {Current-current deformations of conformal field theories, and WZW models}},   {\rm JHEP} {\bf 05} (2003) 071, [\href{http://arxiv.org/abs/hep-th/0304234}{{\tt hep-th/0304234}}].

\bibitem{Kiritsis:1993ju}
E.~Kiritsis, { \it {Exact duality symmetries in CFT and string theory}},   {\rm Nucl. Phys. B} {\bf 405} (1993) 109--142, [\href{http://arxiv.org/abs/hep-th/9302033}{{\tt hep-th/9302033}}].

\bibitem{Gaberdiel:1995mx}
M.~R. Gaberdiel, { \it {Abelian duality in WZW models}},   {\rm Nucl. Phys. B} {\bf 471} (1996) 217--232, [\href{http://arxiv.org/abs/hep-th/9601016}{{\tt hep-th/9601016}}].

\bibitem{Maloney:2020nni}
A.~Maloney and E.~Witten, { \it {Averaging over Narain moduli space}},   {\rm JHEP} {\bf 10} (2020) 187, [\href{http://arxiv.org/abs/2006.04855}{{\tt arXiv:2006.04855}}].

\bibitem{Afkhami-Jeddi:2020ezh}
N.~Afkhami-Jeddi, H.~Cohn, T.~Hartman, and A.~Tajdini, { \it {Free partition functions and an averaged holographic duality}},   {\rm JHEP} {\bf 01} (2021) 130, [\href{http://arxiv.org/abs/2006.04839}{{\tt arXiv:2006.04839}}].

\bibitem{Dong:2021wot}
J.~Dong, T.~Hartman, and Y.~Jiang, { \it {Averaging over moduli in deformed WZW models}},   {\rm JHEP} {\bf 09} (2021) 185, [\href{http://arxiv.org/abs/2105.12594}{{\tt arXiv:2105.12594}}].

\bibitem{Dymarsky:2020qom}
A.~Dymarsky and A.~Shapere, { \it {Quantum stabilizer codes, lattices, and CFTs}},   {\rm JHEP} {\bf 03} (2021) 160, [\href{http://arxiv.org/abs/2009.01244}{{\tt arXiv:2009.01244}}].

\bibitem{luWZWFGK2026}
Y.~L{\"u}, { \it {Localization and Abelianization of Strings on Group Manifolds: The Non-Simply Connected Case}},   {\rm to appear}.

\bibitem{AlekseevSchomerus1999}
A.~Y. Alekseev and V.~Schomerus, { \it {D-branes in the WZW model}},   {\rm Phys. Rev. D} {\bf 60} (1999) 061901, [\href{http://arxiv.org/abs/hep-th/9812193}{{\tt hep-th/9812193}}].

\bibitem{GawedzkiReis2002}
K.~Gawedzki and N.~Reis, { \it {WZW branes and gerbes}},   {\rm Int. J. Mod. Phys. A} {\bf 17} (2002) 5127--5144, [\href{http://arxiv.org/abs/hep-th/0205233}{{\tt hep-th/0205233}}].

\bibitem{FreedHopkinsTelemanI}
D.~S. Freed, M.~J. Hopkins, and C.~Teleman, { \it {Loop groups and twisted K-theory I}},   {\rm J. Topol.} {\bf 4} (2011), no.~4 737--798, [\href{http://arxiv.org/abs/0711.1906}{{\tt arXiv:0711.1906}}].

\bibitem{FreedHopkinsTelemanII}
D.~S. Freed, M.~J. Hopkins, and C.~Teleman, { \it {Loop groups and twisted K-theory II}},   {\rm J. Am. Math. Soc.} {\bf 26} (2013), no.~3 595--644, [\href{http://arxiv.org/abs/math/0511232}{{\tt math/0511232}}].

\bibitem{FreedHopkinsTelemanIII}
D.~S. Freed, M.~J. Hopkins, and C.~Teleman, { \it {Loop groups and twisted K-theory III}},   {\rm Annals Math.} {\bf 174} (2011), no.~2 947--1007, [\href{http://arxiv.org/abs/math/0312155}{{\tt math/0312155}}].

\bibitem{PolyakovWiegmann1984}
A.~M. Polyakov and P.~B. Wiegmann, { \it {Goldstone Fields in Two-Dimensions with Multivalued Actions}},   {\rm Phys. Lett. B} {\bf 141} (1984) 223--228.

\bibitem{Waldorf2008thesisGerbe}
K.~Waldorf,  {\rm {Algebraic Structures for Bundle Gerbes and the Wess-Zumino Term in Conformal Field Theory}}.
\newblock PhD thesis, U. Hamburg (main), 2008.

\bibitem{Alvarez-GaumeGinsparg1984Anomaly}
L.~Alvarez-Gaume and P.~H. Ginsparg, { \it {The Structure of Gauge and Gravitational Anomalies}},   {\rm Annals Phys.} {\bf 161} (1985) 423. [Erratum: Annals Phys. 171, 233 (1986)].

\bibitem{HullSpence1990GaugedWZ}
C.~M. Hull and B.~J. Spence, { \it {The Geometry of the gauged sigma model with Wess-Zumino term}},   {\rm Nucl. Phys. B} {\bf 353} (1991) 379--426.

\bibitem{ManesStoraZumino1985Chiral}
J.~Manes, R.~Stora, and B.~Zumino, { \it {Algebraic Study of Chiral Anomalies}},   {\rm Commun. Math. Phys.} {\bf 102} (1985) 157.

\bibitem{MoraOTZ2006Transgression}
P.~Mora, R.~Olea, R.~Troncoso, and J.~Zanelli, { \it {Transgression forms and extensions of Chern-Simons gauge theories}},   {\rm JHEP} {\bf 02} (2006) 067, [\href{http://arxiv.org/abs/hep-th/0601081}{{\tt hep-th/0601081}}].

\bibitem{Nakahara2003book}
M.~Nakahara,  {\rm {Geometry, topology and physics}}.
\newblock 2003.

\bibitem{MooreSaxena2025TASI}
G.~W. Moore and V.~Saxena, { \it {TASI Lectures On Topological Field Theories And Differential Cohomology}},  in  {\rm {Theoretical Advanced Study Institute in Elementary Particle Physics 2023}: {Aspects of Symmetry}}, 10, 2025.
\newblock \href{http://arxiv.org/abs/2510.07408}{{\tt arXiv:2510.07408}}.
\newblock With an appendix by Daniel S. Freed.

\end{thebibliography}\endgroup
\end{document}